\documentstyle[11pt,aaspp4,psfig]{article}

\authoraddr{C.L. Carilli, NRAO, P.O. Box O, Socorro, NM, 87801, USA}

\authoraddr{M.A. Holdaway, 
NRAO, Campus Building 65, 949 North Cheery Ave., Tucson, AZ
85721, USA }

\slugcomment{To appear in {\it Radio Science}  1999.}

\begin{document}

\title{Tropospheric Phase Calibration in Millimeter Interferometry}
 
\author{C.L. Carilli}
\affil{NRAO, P.O. Box O, Socorro, NM, 87801, USA \\
ccarilli@nrao.edu}

\author{M.A. Holdaway}
\affil{NRAO, Campus Building 65, 949 North Cheery Ave., Tucson, AZ
85721, USA \\
mholdawa@nrao.edu}
 
\begin{abstract}

We review millimeter interferometric phase variations
caused by variations in the precipitable water vapor content of the
troposphere, and we 
discuss techniques proposed to correct for these 
variations. We present observations with the Very
Large Array at 22 GHz and 43 GHz designed to test these techniques.
We find that both the Fast Switching and Paired Array calibration
techniques are effective at  reducing tropospheric phase noise for radio
interferometers. In both cases,
the residual  rms phase fluctuations after correction are {\sl
independent of baseline length for b $>$ b$_{\rm eff}$.} These
techniques allow for diffraction limited
imaging of faint sources on arbitrarily long baselines at mm
wavelengths. 

We consider the technique of tropospheric phase correction 
using a measurement of the  precipitable water vapor content of
the troposphere via a radiometric 
measurement of the brightness temperature of the atmosphere. 
Required sensitivities range
from 20 mK at 90 GHz to 1 K at 185 GHz for the MMA, and 120 mK for the
VLA at 22 GHz. The minimum gain stability
requirement is 200 at 185 GHz at the MMA assuming that the
astronomical receivers are used for radiometry. This increases to 2000
for an uncooled system. The stability requirement is 450 for the
cooled system at the VLA at 22 GHz. 
To perform absolute radiometric phase corrections also 
requires  knowledge of the tropospheric parameters and models
to an accuracy of a few percent.
It may be possible to perform an 
`empirically calibrated' radiometric phase correction, in which 
the relationship between fluctuations in brightness temperature
differences with fluctuations in interferometric phases
is calibrated by observing a strong celestial calibrator at regular
intervals.  A number of questions remain concerning
this technique, including: 
(i) over what time scale and distance will this technique  allow
for radiometric phase corrections when switching between
the source and the calibrator? and
(ii) how often will calibration of the T$_{\rm B}^{\rm rms}$ - $\phi_{rms}$
relationship be required?


\end{abstract}



\section{Introduction}

The most important difference between mm and cm
interferometry is the effect of the troposphere at mm wavelengths.
The optical depth of the troposphere becomes significant below 1 cm,
leading to increased system temperatures due to atmospheric emission,
and increased demands on gain calibration due to variable
opacity (Yun et al., 1998, Kutner and Ulich 1981). Even more dramatic
is the effect of the troposphere on 
interferometer phases. Variations in the tropospheric water vapor column
density lead to variations in electronic pathlength, and hence variations in
interferometric phase. This can cause loss of amplitude of the cross
correlations, or `visibilities', 
over the integration time (`coherence'), reduced spatial resolution
(`seeing'), and pointing errors (`anomalous refraction').

Until recently, mm interferometric arrays have been restricted to
maximum baselines of a few hundred meters. Diffraction limited
resolution images at mm wavelengths can be made on such baselines,
since the tropospheric phase fluctuations on such baselines at good
observatory sites are typically one radian or less under good weather
conditions. However, many existing mm arrays are expanding to
baselines of 1 km or more, and the planned large millimeter array
facility at the Chajnantor site in Chile will have baselines as long
at 10 km. Even at this premium site, phase fluctuations due to the
troposphere will be larger than 1 radian at 230 GHz on these
baselines, except under the best weather conditions. Hence,
tropospheric seeing will preclude diffraction limited resolution
imaging at frequencies of 230 GHz for arrays larger than about 1 km,
even at the best possible sites, if no corrections are made for
tropospheric phase noise.  At higher frequencies, the maximum
baselines which would permit uncorrected observations would be even
shorter.

In this paper we review the theory of tropospheric phase noise
in mm interferometry, along with examples showing tropospheric induced
phase fluctuations and their effect on
images made with interferometric arrays. 
We then consider three techniques for reducing tropospheric phase
noise: (i) Fast Switching phase calibration, (ii) Paired Array phase
calibration, and (iii) radiometric phase calibration. 
The first two techniques entail using celestial calibration sources
near the target source, with either a calibration cycle time fast
enough to `stop' the troposphere (Fast Switching), or by using some of
the antennas as a tropospheric `calibration array' (Paired Array). 
We present extensive observational data 
using the Very Large Array (VLA) in Socorro, NM, USA, 
at 22 GHz and 43 GHz  designed to
test the efficacy of these first two techniques on baselines longer
than a few km. 

The radiometric phase correction technique entails real-time
estimation of the precipitable water vapor content along each
antenna's line of sight through the troposphere via a radiometric
measurement of the brightness temperature of the atmosphere above each
antenna.  A number of issues are addressed, including: (i) the
required radiometric sensitivity as a function of frequency and site
quality, (ii) the constraints on ancillary data, such as atmospheric
data and models, in order to perform an absolute radiometric phase
correction, and (iii) the limitations to making radiometric phase
corrections by calibrating the relationship between brightness
temperature fluctuations and interferometric phase using celestial
sources.

\section{A General Description of the Troposphere and 
the Mean Tropospheric Effect on Interferometric Phase}

The troposphere is the lowest layer of the atmosphere, extending
from the ground to the stratosphere at an elevation of 7 km to 10 km.
The temperature decreases with altitude in this layer, clouds form,
and  convection can be significant (Garratt 1992).
The troposphere is composed predominantly of N$_2$, O$_2$, plus
trace gases such as water vapor, N$_2$O, and CO$_2$, plus
particulates such as liquid water and dust in clouds.
The troposphere becomes increasingly opaque with increasing frequency,
mostly due to absorption by  O$_2$ and H$_2$O.

Figure 1 shows models of the atmospheric transmission
at cm and mm wavelengths for the VLA site at 2150
m altitude, and the planned
millimeter array (MMA)  site in Chile at 4600 m altitude
(Holdaway and Pardo 1997, Liebe 1989).
The plot shows a series of strong absorption lines including the water
lines at 22 GHz and 183 GHz, and the
O$_2$ lines at 60 GHz and 118 GHz, plus a systematic decrease in the
transmission with increasing frequency between the lines. This
`pseudo-continuum' opacity is due to the sum of the pressure broadened
line wings of a multitude of sub-mm and IR lines of  water vapor.
The  plot for the MMA site at Chajnantor in Chile
includes the typical value for the column density of
precipitable water vapor, w$_{\circ}$ = 1 mm, while the water vapor
column for the VLA site is assumed to be
w$_{\circ}$ = 4 mm, where precipitable water vapor (PWV) = the depth of the
water vapor if converted to the liquid phase.
Figure 2 shows the relative contributions from water vapor and 
dry air (O$_2$ plus other trace gases) for the VLA site. Below 130 GHz,
both O$_2$ 
and H$_2$O contribute significantly to the optical depth. Above 130
GHz, H$_2$O dominates the optical depth. 

The troposphere has a non-unit refractive index, $n$.
The refractive index is defined  by the phase change experienced
by an electromagnetic wave, $\phi_e$, propagating over a physical
distance, D:
$$ \phi_e = {{2 \pi}\over{\lambda}} \times n \times \rm D, $$
or, in terms of `electrical pathlength', L$_{\rm e}$:
$$  \rm L_{e} = \lambda \times {{\phi_e}\over{2 \pi}} = {\it n}  \times D $$
The refractive index of air is non-dispersive
(ie, independent of frequency) except near the strong resonant
water and O$_2$ lines, and is  typically given as a difference
with respect to vacuum ($n_{vacuum}$ $\equiv$ 1),
in parts per million, $N$, as (Waters 1976):
$$ N \equiv (n - 1) \times 10^{6} $$
The index of refraction of air is typically separated into
the dry air component, $N_{d}$, and the water vapor component,
$N_{wv}$. These terms behave as (Waters 1976, Bean and Dutton
1968):
\begin{eqnarray}
N_{d} = 2.2 \times 10^{5} \times \rho_{tot}  \nonumber  \\
N_{wv} = 1.7 \times 10^{9} \times {{\rho_{wv}}\over{\rm T_{\rm atm}}}
\nonumber
\end{eqnarray}
where $\rho_{tot}$ is the total mass density in gm cm$^{-3}$, 
and $\rho_{wv}$ is the water vapor  mass density.
The inverse 
dependence on temperature for $N_{wv}$ is due to the
increased effect of collisions on the (mis-)alignment of the permanent
electric dipole moments of the water molecules with increasing
temperature (Waters 1976, Bean and Dutton 1968).
A detailed derivation of these relationships can be found in 
Thompson, Moran, and Swenson (1986).


For water vapor alone, it can be shown that $\rho_{wv}$ = ${\rm
w}\over{\rm D}$. Using the equations above then leads to the 
relationship between the electrical pathlength, L$_{\rm e}$,  and the
precipitable water  vapor column, w:
\begin{eqnarray}
\rm L_e = 1.7\times10^3 {{w}\over{T_{\rm atm}}} \approx 6.3\times\rm w
\nonumber 
\end{eqnarray}
or:
\begin{equation}
\phi_e \approx {{12.6 \pi}\over{\lambda}} \times \rm w 
\end{equation}
for T$_{\rm atm}$ $\approx$ 270 K. This relation between electrical
pathlength and precipitable water  vapor column
has been verified experimentally for a number of atmospheric
conditions (Hogg, Guiraud, and Decker 1981).

\section{Phase Variations due to the Troposphere}

Variations in precipitable water vapor lead to
variations in the effective electrical path length, corresponding
to variations in the phase of an electromagnetic wave
propagating through the troposphere (Tatarskii 1978).
Such variations are seen as
`phase noise' by radio interferometers. Since the troposphere is
non-dispersive, the phase contribution by a given amount of water vapor
increases linearly with frequency (except in the vicinity of
the strong water lines). Hence,  tropospheric phase
variations are most prominent for mm and
sub-mm interferometers, and can be the limiting factor for
the coherence time and spatial resolution of mm interferometers
(Hinder and Ryle 1971, Lay 1997, Wright 1996).

The standard model for tropospheric phase
fluctuations involves  variations in the water vapor column density
in a turbulent layer in the troposphere with
a mean height, h$_{\rm turb}$, and a vertical extent, W, which
moves at some velocity, v$_a$.
This model includes the `Taylor hypothesis', or `frozen screen
approximation', which states that: `if the turbulent intensity
is low and the turbulence is approximately stationary and homogeneous,
then the turbulent field is unchanged over the atmospheric boundary
layer time scales of interest and advected with the mean wind'
(Taylor 1938, Garratt 1992).
Under this assumption one can relate temporal and spatial
phase fluctuations  with a simple Eulerian transformation
between baseline length, b,  and v$_a$:~ b = v$_a$$\times$time. 
In the following sections we adopt a value of
v$_a$ = 10 m s$^{-1}$. This process is shown schematically in Figure
3.

A demonstration of tropospheric phase fluctuations is shown in Figure
4 for observations made with the VLA at 22 GHz. The VLA is an
aperture synthesis array comprised of 27 parabolic
antennas of 25m diameter, operating at frequencies from 75 MHz to 50
GHz (Napier, Thompson, and Ekers 1983). The antennas are situated in a
`Y' pattern, along three arms situated north,  southwest, and
southeast. The maximum physical  baseline in the largest configuration
is 33 km. The complex cross correlations of the electric fields
measured at each antenna are calculated between all pairs of antennas
(`interferometers') as a function of time. 
The Fourier transform of
these measurements of the spatial coherence function of the electric
field  then gives the sky brightness distribution (Clark 1998). 

For these observations, two
subarrays were employed, one observing the celestial calibrator
0423+418, and the second observing the calibrator 0432+416. 
The sub-arrays were `inter-laced', meaning that every second antenna
along each arm of the array observed a given source. 
Antenna-based phase and amplitude solutions were derived from the data
in each sub-array using self-calibration 
with an averaging time of 30 sec (Cornwell 1998).
Figure 4 shows the
antenna-based phase solutions from two pairs 
of neighboring antennas along the southwest arm. The antennas at
stations W16 and W4 were 
observing 0423+418 while the antennas at W18 and W6 were observing
0432+416. For adjacent pairs of antennas (W16-W18 and W6-W4) the
temporal variations in the phase track each other closely. 
This close relationship for phase variations 
between neighboring antennas in the two different
subarrays  is the signature that the phase
variations are primarily tropospheric in origin, and are correlated on
relevant timescales and baseline lengths. 

An important aspect of tropospheric phase fluctuations arising from
the Taylor hypothesis is the relationship between the amplitude of the
fluctuations and the time scale: large amplitude fluctuations occur
over long periods and are partially correlated between antennas, 
while small amplitude fluctuations occur over short periods and are
uncorrelated between antennas, depending on the baseline length. 
This effect can be seen in Figure 4 by the fact that the antennas at
the outer stations (W14 and W16) show larger amplitude fluctuations
relative to the inner stations (W4 and W6). This occurs because the
antenna-based phase solutions for each subarray are referenced to
antennas at the 
center of the array, such that the reference antennas are within about
250 m of W4 and W6, but are separated from W16 and W14 by about 
2000 m.

An example of what occurs in the image plane due to tropospheric phase
fluctuations is shown in Figure 5. Observations were made of the
celestial calibrator 2007+404 at 22 GHz with the VLA at a resolution
of 0.1$''$ (maximum baseline = 30 km) for a period of 1 hour.
The data were self-calibrated using a long solution averaging time of
30 minutes, ie. just a mean phase was removed from each half of the
data. `Snap-shot' images were then made from one minute of data at the
beginning and end of the observation (upper left and upper right
frames in Figure 5, respectively). Two important trends are apparent
in these two frames. First, notice the positive-negative
side-lobe pairs straddling the peak, indicative of antenna-based
phase errors (Ekers 1998). These image artifacts are due to phase
fluctuations which arise in small scale water vapor structures in the
troposphere that are not correlated between antennas. Second, notice
that the peak in each image has shifted from the true source  position.
This position shift is due to phase fluctuations
which arise in large scale water vapor structures in the troposphere
that are correlated between antennas, ie. a phase gradient across the
array. These two frames are analogous to optical `speckle' images,
although the timescales in the radio are much longer than in the
optical due to the larger spatial scales for the turbulence. 

The lower left frame shows the image of 2007+404 made using the
full hour of data, but with a self-calibration averaging time of 
30 minutes. The source appears extended in this image. The lower right
frame shows the same image after self-calibration with an averaging
time of 30 seconds, and in this case the source is
unresolved, ie. the source size is equivalent, within the noise, to
the interferometric synthesized beam.
The lower left image has a peak surface brightness of 1.0
Jy beam$^{-1}$, an off-source rms noise level of 47 mJy beam$^{-1}$,
and a total flux density of 1.5 Jy. The lower right 
image has a peak surface brightness of 1.6
Jy beam$^{-1}$, an off-source rms noise level of 5 mJy beam$^{-1}$,
and a total flux density of 1.6 Jy.
Not correcting for tropospheric phase noise in the lower left frame
has: (i) increased the off-source
noise in the image, (ii) decreased the `coherence' (ie. lowered
the peak surface brightness), 
and (iii) degraded the resolution (`seeing').

The important lesson from Figure 5 is that, while tropospheric phase
errors can be quantified in terms of an antenna-based phase error, the
errors are partially correlated between antennas on certain
spatial and temporal scales, leading to positional shifts of sources
as well as the standard positive-negative side-lobe pairs. 
Also, it is important to keep in mind that short baselines
only `sample' the power in the phase screen on scales of order
the baseline length.


\section{Root Phase Structure Function}

Tropospheric phase fluctuations are usually
characterized by the spatial phase structure function, $\rm D_{\Phi}(b)$,
\begin{equation}
\rm D_{\Phi}(b) \equiv \langle ( \Phi(x + b) - \Phi(x) )^2 \rangle,
\end{equation}
where b is the distance between two antennas, $\rm \Phi(x + b)$ is the
atmospheric phase measured at one antenna and $\rm \Phi(x)$ is the
atmospheric phase measured at the second antenna, and the brackets
represent an ensemble average.  Usually in radio astronomy the
ensemble average is replaced by a time average on one particular
baseline.  An interferometric array will sample the phase structure
function at several baselines.  For a single interferometer, the
Taylor hypothesis (which asserts that temporal phase fluctuations are
equivalent to spatial phase fluctuations) permits us to measure
temporal phase fluctuations on a single baseline and translate these
into the equivalent spatial phase structure function.  In the
following discussion we consider the square root of the phase
structure function (the `root phase structure function'), which
corresponds to the rms phase variations as a function of baseline
length:
$$\rm \Phi_{rms} \equiv \sqrt{D_{\Phi}}$$.

Kolmogorov turbulence theory  (Coulman 1990) predicts a  function of
the form:
\begin{equation}
\rm \Phi_{\rm rms}(b) = {K\over{\lambda_{mm}}} ~b^{\alpha}
~~ deg,
\end{equation}
where b is in km, and $\lambda$ is in mm. A
typical value of K = 100 for the MMA site in Chajnantor under
good weather conditions, and K = 300 is typical
for the VLA site (Carilli, Holdaway, and Sowinski 1996, Sramek 1990).

Kolmogorov turbulence theory predicts
$\alpha$ = ${1\over3}$ for baselines longer than the width of the
turbulent layer, W, and $\alpha$ = ${5\over6}$ for baselines
shorter than W (Coulman 1990). The change in power-law index at  b $=$
W  is due to the finite vertical extent
of the turbulent  layer. For baselines
shorter than W the full 3-dimensionality
of the turbulence is involved (thick-screen), while for longer baselines
a 2-dimensional  approximation applies (thin-screen).
Turbulence theory also predicts an `outer-scale', L$_{\circ}$, beyond which the
rms phase variations should not increase with baseline length
(ie. $\alpha$ = 0). This scale corresponds to
the largest coherent structures, or maximum correlation length,
for water vapor fluctuations in the troposphere, presumably set by
external boundary conditions.

Recent observations with the VLA 
support Kolmogorov theory for tropospheric phase fluctuations.
Figure 6 shows the root phase structure function made using the BnA
configuration of the VLA. This configuration has good baseline
coverage ranging from 200m to 20 km, hence sampling all three
hypothesized ranges in the structure function.
Observations were made at 22 GHz of 
the VLA calibration source 0748+240. The total observing time
was 90 min, corresponding to a tropospheric travel distance of 54 km,
using v$_a$ = 10 m s$^{-1}$ (see section 6.2).
The open circles show the nominal tropospheric root phase structure
function over the full 90 min time range.\footnote{Note that the 
total observing time for calculating the rms phase fluctuations
must be long for the larger configurations of the
VLA, since the phase variations on a 
given baseline may have a significant, and perhaps even dominant,
contribution from structures in the troposphere as large as five times
the baseline length (Lay 1997).}
The solid squares are the rms phases after subtracting (in
quadrature) a constant electronic noise term of 10$^{\circ}$, as derived from
the data by requiring the best power-law on short baselines.
The 10$^{\circ}$ noise term is consistent with previous measurements at the
VLA indicating electronic phase noise increasing with frequency as
0.5$^{\circ}$ per GHz (Carilli and Holdaway 1996).

The three regimes of the structure function as predicted
by Kolmogorov theory are verified in Figure 6.
On short baselines (b $\le$ 1.2 km) the measured power-law index
is 0.85$\pm$0.03 and the predicted value is 0.83. On intermediate
baselines (1.2 $\le$ b $\le$ 6 km)
the measured index is 0.41$\pm$ 0.03 and the predicted value
is 0.33. On long baselines (b $\ge$ 6 km) the measured index is
0.1$\pm$0.2 and the predicted value is zero. The implication is that
the vertical extent of the turbulent  layer is:~ W
$\approx$ 1 km, and that the outer scale of the turbulence is:~
L$_{\circ}$ $\approx$ 6 km.  The increase in the scatter of the rms phases
for baselines longer than
6 km  may be due to an anisotropic outer scale (Carilli and
Holdaway 1997).

In practice, single baseline 
site testing interferometers tend to see values of $\alpha$ which form a
continuous distribution between the theoretical thin and thick layer
values of ${1\over3}$ and ${5\over6}$ 
(Holdaway et al., 1995). The distribution cuts off
fairly sharply at these two theoretical values.  Such a distribution
could be due to the presence of both a thin turbulent layer associated
with the ground or an inversion layer and a thick turbulent layer.  By
changing the relative weight of each of these two layers with their
theoretical power law exponents, the resulting phase structure
function will be very close to a power law with an exponent between
the two theoretical values.  Alternatively, the intermediate power law values
could just reflect the transition between thin and thick turbulence.

\section{Effects of Tropospheric Phase Noise}

\subsection{Coherence} 

Tropospheric phase noise leads to a number of adverse effects on
interferometric observations at mm wavelengths. First is the loss of
coherence of a measured visibility on a given baseline over a given
averaging time due to phase variations.  For a given visibility, V =
V$_{\circ}$e$^{i \phi}$, the effect on the measured amplitude due to phase
noise in a given averaging in time is:
\begin{equation}
<\rm V> = \rm V_{\circ} \times <e^{i \phi}> = V_{\circ} \times e^{-\phi_{\rm rms}^2/2}
\end{equation}
assuming Gaussian random phase fluctuations with an
rms variation of $\phi_{\rm rms}$ over the averaging time (Thompson,
Moran, and Swenson 1986). For example,
for $\phi_{\rm rms}$ = 1 rad, the coherence is:~ ${<\rm V>}\over{\rm
V_{\circ}}$ = 0.60, meaning the observed visibility amplitude is reduced by
40$\%$ from the true value.

\subsection{Seeing} 

A second effect of tropospheric phase
fluctuations is to
limit the spatial resolution of an observation in a manner analogous to
optical seeing, where optical seeing is due to thermal
fluctuations rather than water vapor fluctuations. 
Since interferometric phase corresponds to the
measurement of the position of a point
source (Perley 1998), it is clear that
phase variations due to the troposphere will lead to positional
variations of a source, and hence `smear-out' a point source image
over time (Figure 5). The magnitude of tropospheric seeing can be
calculated by considering the coherence as a function of baseline
length. Since the coherence decreases for longer baselines given an
averaging time long compared to the array crossing time, the observed
visibility amplitude decreases with increasing baseline length, as
would occur if the source were resolved by the array. Using equation 3
for the root phase  structure, and equation 4 for the coherence, the
visibility amplitude as a function of baseline length becomes:
\begin{equation}
<\rm V > = V_{\circ} \times exp(-[{{K'  b^\alpha}\over{\lambda \sqrt{2}}}]^2)
\end{equation}
Note that the exponent must be in radians, so K$'$ = K $\times$  ${2
\pi}\over{360}$.
The baseline length corresponding to the half-power point of the
visibility curve, b$_{1/2}$, then becomes:
$$
\rm b_{1/2} = (1.2 \times {{{\lambda_{mm}}\over{K'}}})^{1/\alpha} ~~km
$$

For example, at 230 GHz  using
the typical value of $\alpha$ = 5/6, and a typical value for K$'$ at
the MMA site of 1.7, the value of b$_{1/2}$ =  0.9 km.
This means that the resolution of the array is limited by tropospheric
seeing to: $\theta_{seeing}$ 
$\approx$ ${\lambda}\over{b_{1/2}}$ $\approx$ 0.3$''$ at 230 GHz.
For average weather conditions, tropospheric seeing precludes
diffraction limited resolution imaging 
for arrays larger than about 1 km at the MMA site in Chile,
if no corrections are made for tropospheric phase noise. 
A rigorous treatment of tropospheric seeing, with predicted
source sizes under various assumptions about the turbulence, can be
found in Thompson, Moran, and Swenson (1986).

Two important points need to be remembered when considering
tropospheric seeing. First is that the root phase structure function
flattens dramatically on baselines longer than $\approx$ 1 km, such
that the tropospheric seeing degrades slowly with longer baselines.
And second, there is an
explicit connection between the coherence loss and the seeing: on the
short baselines, where the phase errors are smaller, there is less
coherence loss, and the correct flux density is measured even for a long
averaging time.  On the longer
baselines, the phase errors are larger, causing decorrelation of the
visibilities, either within an integration or across many
integrations, thereby fictitiously resolving the source.  It is the
selective loss of coherence on the long baselines which determines the
seeing.
This phenomenon can be seen in the
lower left frame of Figure 5, in which the peak surface brightness is
only 60$\%$ of the expected peak, but the total flux density averaged
over the `seeing disk' is 94$\%$ of the true value, ie. the shortest
baselines see the total flux density of the source even for long
averaging times.

\subsection{Anomalous Refraction}

A final problem arising from tropospheric phase variations is
`anomalous refraction', or tropospheric induced pointing errors
(Holdaway 1997, Butler 1997, Holdaway and Woody 1998).  This effect
corresponds to tropospheric seeing on the scale of the antenna
itself. Phase gradients across the antenna change the apparent
position of the source on a time scale $\approx$ ${D}\over{\rm v_a}$
$\approx$ 1 second, for an antenna with a diameter $D = 10$ m.  A
straight-forward application of Snells' law shows that the effect in
arc-seconds should decrease with antenna diameter as $D^{\alpha - 1}$,
or about $D^{-0.4}$ for median $\alpha$ of 0.6.  The decrease in the
pointing error with dish diameter is due to the fact that the value of
$\alpha$ in the root phase structure function is less than unity, and
hence the angle of the `wedge' of water vapor across the antenna
becomes shallower with increasing antenna size. However, in terms of
fractional beam size, the effect becomes worse with antenna size as
$D^{\alpha}$. For the 10m MMA antennas the expected magnitude of the
effect at an elevation of 50$^o$ is $\approx$ 0.6$''$, which is about
50$\%$ of the pointing error budget for the antennas.

\section{Stopping the Troposphere: Techniques to Reduce the
Effects of Tropospheric Phase Noise}

An important point to keep in mind is that while tropospheric phase
variations can be quantified in terms of a baseline-length
dependent structure function, the errors are fundamentally
antenna-based, and hence can be corrected by antenna-based calibration
schemes, such as  self-calibration or fast switching calibration.

\subsection{Self-Calibration}

A straight forward method of reducing phase errors due to the
troposphere is self-calibration (Cornwell 1998). Self-calibration
removes the baseline-dependent term in the root structure function,
$\Phi_{\rm rms}$(b), leaving the residual tropospheric phase noise
dictated by the `effective baseline': b$_{\rm eff}$ = ${\rm v_a
t_{\rm ave}}\over2$ 
= half the distance the troposphere moves during the self-calibration 
averaging time, t$_{\rm ave}$. The factor of two arises from the fact that
the mean calibration applies to the middle  of the solution interval.
The Taylor hypothesis dictates a 
relationship between temporal and spatial fluctuations
such  that the longer baselines
will not sample the full power in the root phase structure function if
the calibration cycle time is shorter than the baseline crossing time
for the troposphere.

Of course, we would like to make $\rm t_{\rm ave}$ as short as possible, but
for a target source of some given brightness, we are limited in that we
must detect the source in t$_{\rm ave}$ on each baseline with sufficient
signal-to-noise ratio (SNR $\approx$ 2 for arrays with large numbers
of antennas; Cornwell 1998) to be 
able to solve for the phase.  Hence, there will be sources which are so
weak that they cannot be detected in a time short enough to track the
atmospheric phase fluctuations.
For the MMA at
230 GHz, self-calibration should be possible on fairly weak  continuum
sources (of order 10 mJy), with fairly short integration times
($\approx$ 30 sec), leading to  residual rms phase errors $\le$
20$^o$. For the completed VLA  43 GHz system the limit is 50 mJy
sources with 30 second averaging times with residual rms phase
variations of 10$^o$. Self-calibration is not possible for weaker
continuum sources, or for weak spectral line
sources,  or in the case where absolute positions are required. In
these cases  other
methods must be employed to `stop' tropospheric phase variations.

\subsection{Fast Switching}

Another method for reducing tropospheric phase variations is
`Fast Switching' (FS) phase calibration. This method is simply normal
phase calibration using celestial calibration sources close to
the target source (Fomalont and Perley 1998), only with a 
calibration cycle time, t$_{cyc}$, short enough to reduce tropospheric
phase variations to an acceptable level (Holdaway 1992, Holdaway and
Owen 1995, Carilli and Holdaway 1996, 1997).
The expected residual phase fluctuations after FS calibration
can be derived from the root phase structure function (equation 3),
assuming an `effective baseline length', b$_{\rm eff}$, given by:
\begin{equation}
\rm b_{\rm eff}~ \approx~ d~ +~ {{v_a t_{cyc}}\over2}
\end{equation}
where $v_a$ = wind speed, and d = the physical
distance in the troposphere between the calibrator and source. 
The FS technique will be effective for calibration cycle times
shorter than the baseline crossing time of the troposphere = ${\rm
b}\over{\rm v_{a}}$.
Moreover, a significant gain is made when b$_{\rm eff}$ $<$ 1 km, thereby
allowing for corrections to be made on the steep part of the root
phase structure function (Figure 6), implying a timescale of 200
seconds or less for effective FS corrections.
As with self-calibration, the calibrator source must be detected with
sufficient SNR on time scales short enough to track the
atmospheric phase fluctuations. 

The effectiveness of FS phase calibration is shown
in Figure 7 for 22 GHz data from the VLA on
baselines ranging from 100 m to 20 km.
The solid squares show the nominal tropospheric root phase
structure function averaged over 90 minutes (Figure 6).
The open circles are the rms phases of the visibilities after applying
antenna based phase solutions averaged over 300 seconds. The stars are the
rms phases of the visibilities after applying
antenna based phase solutions averaged over 20 seconds.
The residual root structure function using a 300 second calibration cycle
parallels the nominal tropospheric root structure function out to a
baseline length of 1500m, beyond which the root structure function
saturates at a constant rms phase value of
20$^{\circ}$. The implied wind velocity is then:~ v$_a$ = ${2 \times
1500 m}\over{300 sec}$ = 10 m s$^{-1}$. Using a 20 second calibration
cycle reduces b$_{\rm eff}$ to only 100 m, which is shorter than  the shortest
baseline of the array, and the saturation rms is 5$^{\circ}$.

The important point is that, after applying standard phase calibration
techniques on timescales short compared the array crossing time of the
troposphere, the resulting rms phase fluctuations are {\sl independent
of baseline length for b $>$ b$_{\rm eff}$.} The FS technique
allows for diffraction limited
imaging of faint sources on arbitrarily long baselines. Note that this 
conclusion should also apply to Very Long Baseline Interferometry (VLBI), 
although the problems are accentuated due to the fact that 
antennas can be observing at very different elevations at 
any given time, and that VLBI observations typically employ low
elevation observations in order to maximize mutual visibility times for
widely separated antennas (see the discussion of the 
VLBI phase referencing technique by Beasley and Conway 1995). 

An important question to address when considering FS phase
calibration is: are there  enough calibrators in the
sky in order to take advantage of a switching time as short as 40
seconds? This depends on the slew rate and settling time of the
telescope, the  set-up time of the electronics, the sensitivity of the
array, and the sky surface density of celestial calibrators.
Holdaway, Owen, and Rupen (1994) estimated the calibrator
source counts at 90~GHz.  They measured the 90~GHz flux densities of 367
flat spectrum quasars known from centimeter wavelength surveys, thereby
determining the distribution of the spectral index between 5~GHz and
90~GHz, which was statistically 
independent of source flux density.  By applying this spectral index distribution to
well understood 5~GHz flat spectrum source counts, they were able to
estimate the integral source counts for appropriate phase calibrators
at 90~GHz. They estimate that the integral source counts over the
whole sky between 0.1 to 1.0 Jy is about 170 $S^{3/2}_{90}$, where
S$_{90}$ is the 90 GHz source flux 
density in Jy.  Then, the typical distance to
the nearest appropriate calibrator at 90~GHz will be about 
$\rm \theta \approx 7 \times S_{\nu}^{0.75}~~ deg$.
For calibrators with S$_{\nu}$
$\ge$ 50 mJy, and assuming  a slew rate of 2 deg s$^{-1}$,
the 40 element MMA should be able to employ FS phase calibration on most
sources with total cycle times $\le$ 20 sec, leading to residual rms
phase fluctuations $\le$ 20$^o$ at 230 GHz, and on-source duty cycles
$\approx$ 75$\%$.  One important practical problem is the lack of
all-sky surveys at high frequency from which to generate calibrator
source lists.  However, the MMA will be both agile and sensitive enough to
survey a few square degrees around the target source in a few minutes
prior to the observations, thereby permitting the determination
of the optimal calibrator.

\subsection{Paired Array Calibration}

A third method for reducing tropospheric phase noise for faint
sources is paired antenna, or paired array, calibration.
Paired Array (PA) calibration involves phase calibration of
a `target' array of antennas using a separate `calibration' array,
where the target array is observing continuously a weak source of scientific
interest while the calibration array is observing a nearby calibrator
source (Holdaway 1992, Counselman et al., 1974, Asaki et al., 1996, 1998,
Drashkik and  Finkelstein 1979). In its simplest form
PA calibration implies applying the phase solutions
from  a calibration array antenna
to the nearest target array antenna at each integration time.
An improvement can be made
by interpolating the solutions from a number of nearby calibration
array antennas to a given target array antenna at each integration
time.  Ultimately, the discrete
measurements in space and time of the phases from the
calibration array could be incorporated into a physical model for the
troposphere to solve for, and remove, the effects of the tropospheric phase
screen on the target source as a function of time and space using some
intelligent method of data interpolation, such as forward 
projection using a physical model for the troposphere and
Kalman filtering of the spatial time series (Zheng
1985). 

For simple pairs of antennas, the residual phase error
can be derived from the root phase structure function with:~
$$\rm b_{\rm eff} \approx d + \Delta b$$
where d is the same as in equation 10, and $\Delta$b is the baseline
length between the calibration antenna and the target source antenna.

Figure 4 shows an observation for which PA calibration was
implemented. Again, observations were made at 22 GHz using two
`inter-laced' subarrays observing two close calibrators, 0432+416 and
0423+416. Notice how the phase variations for adjacent antennas in the
different subarrays track each other closely. This correlation between
phase variations from neighboring antennas in different subarrays
observing different sources implies that the tropospheric phase
variations can be corrected using PA calibration. 

In Figure 4 the temporal variations for neighboring antennas track
each other  well, but the mean phase over the observing time range is
different between antennas. This phase off-set  is due
to the electronics and/or optics at each antenna, and should be slowly
varying in time. Before interpolating phase solutions from the
calibration array to the target array one must first determine, and
remove, the electronic phase off-sets. This can be done by observing a
celestial calibrator every 30 min or so. A demonstration of this
process is shown in Figure 8. The upper frame in Figure 8 shows a
random phase distribution along the west arm at a given time
before the mean electronic phase is removed,  while the lower frame in
Figure 8 shows  a smooth phase gradient along the arm after correction
for the electronic phase term, indicating a tropospheric 
phase `wedge' down the array arm at this time. 

A quantitative measure of the effects of PA  calibration
can be seen in the root phase structure function plotted in Figure
9. The open triangles show the 
root phase structure function  for the given observing day, as
determined from the  data with only the mean phase calibration (30 min
averaging) applied. This function can be fit by a power-law in
rms phase versus baseline length with index 0.65.
The rms magnitude of 35$^o$ on  a baseline of 1000m is
somewhat higher than the expected value of about 25$^o$ on a typical
summer evening at the VLA (Carilli etal. 1996).
The stars in Figure 9 correspond to the `noise floor' for
the phase measurements, as determined by calculating the
root structure function from self-calibrated data with an averaging
time of 30 seconds.

The solid squares in Figure 9 show the root structure
function  for the data with PA calibration applied.
The PA process in this case entailed interpolating phase solutions
from neighboring antennas observing the `calibration source' 0423+418,
to the antennas observing the `target source' 0432+416.
The residual rms phase values are about  10$^o$ on short baselines,
and increase very slowly with baseline length. These data indicate
a significant improvement in rms phase fluctuations after
application of PA calibration for
baselines longer than about 300 m.

The increased noise floor for the PA calibrated data
relative to self-calibration indicates residual short-timescale
phase differences which do not replicate between the target and
calibration arrays. 
This noise floor is a combination of 
`jitter' in the electronic phase contribution, and 
residual tropospheric phase noise as determined by b$_{\rm eff}$
above. Note that the residual noise floor increases slowly with
baseline length.  This is due to the logarithmically increasing
separation between VLA antennas along the arm.

\section{Radiometry}

The brightness temperature of atmospheric emission, 
T$_{\rm B}^{\rm atm}$, can be measured  
using a radiometer, and is given by the 
radiometry equation (Dicke et al., 1946):
\begin{equation}
\rm T_{\rm B}^{\rm atm}~ =~ \rm T_{\rm atm} ~ \times ~ (1~ -~
e^{-\tau_{\rm tot}})
\end{equation}
where T$_{\rm atm}$ is the physical temperature of the atmosphere, and 
$\tau_{\rm tot}$ is the optical depth, which depends on, among other things,
the precipitable water vapor content  (PWV) of the troposphere. 
By measuring fluctuations in atmospheric brightness temperature
with a radiometer, one can infer the fluctuations in the column
density of water vapor of the troposphere (Barrett and Chung 1962,
Staguhn et al., 1998, Staelin 1966, Westwater and Guiraud 1980,
Rosenkranz 1989, Bagri 1994, Sutton and Hueckstadt 1997,
Lay 1998). The relationship between electrical pathlength and water
vapor column (equation 1) can then be used to derive the variable
contribution from  water vapor to the interferometric phase (Elgered
1993). This technique has been used with varying degrees of success
at connected-element mm interferometers (Welch 1994, Woody and Marvel
1998,  Bremer et al., 1997), and in geodetic VLBI experiments (Elgered et
al. 1991). 

We assume that the atmospheric opacity can be divided into three parts:
\begin{equation}
\rm \tau_{\rm tot} ~ = ~ A_{\nu}\times w_{\circ}~ +~ B_{\nu}~ +
~A_{\nu}\times w_{\rm rms} 
\end{equation}
where: (i)  A$_{\nu}$ is the
optical depth per mm of PWV as
a function of frequency, 
(ii) w$_{\circ}$ is the temporally stable (mean)
value for PWV of the troposphere, (iii) B$_\nu$ is the total
optical depth due to dry air as
a function of frequency (also assumed to be
temporally stable), and (iv)  w$_{\rm rms}$ is the time variable component
of the PWV of the troposphere. It is this time variable component which
causes the tropospheric phase `noise' for an interferometer.
In effect, we assume a constant mean optical depth: $\rm \tau_{\circ} \equiv
A_{\nu}\times w_{\circ}~ +~ B_{\nu}$, with a fluctuating term due to changes
in PWV:~ $\rm \tau_{\rm rms} \equiv A_{\nu}\times w_{\rm rms}$, and that
$\tau_{\circ}$ $>>$ $\tau_{\rm rms}$.

Inserting equation 8 into equation 7, and making the reasonable assumption
that A$_\nu$$\times$w$_{\rm rms} << 1$, leads to:
\begin{equation}
\rm T_{\rm B} ~ =~ T_{\rm atm}\times [1 - e^{-\tau_{\circ}}] ~~ + ~~
T_{\rm atm}\times e^{-\tau_{\circ}}\times[A_{\nu}\times w_{\rm rms} ~ + ~
{{(A_{\nu}\times w_{\rm rms})^2}\over{2}} ~ + ~ ...]
\end{equation}
The first term on the right-hand side of equation 9 represents the
mean, non-varying T$_{\rm B}$ of the troposphere. The second term represents
the fluctuating component due to variations in PWV, which we define
as:
\begin{equation}
\rm T_{\rm B}^{\rm rms} ~ \equiv ~ T_{\rm atm}\times e^{-\tau_{\circ}}\times[A_{\nu}\times
w_{\rm rms} ~ + ~  {{(A_{\nu}\times w_{\rm rms})^2}\over{2}} ~ + ~
...]
\end{equation}

At first inspection, it would appear that equation 10 applies to
fluctuations in a turbulent layer at the top of the
troposphere, since the fluctuating component is fully attenuated
(ie. multiplied by  $e^{-\tau_{\circ}}$). However, for a turbulent
layer at lower altitudes there is the additional term of attenuation
of the atmosphere above the turbulent layer by the turbulence.
It can be shown that the terms exactly cancel for an isobaric,
isothermal atmosphere, in which case
equation 10  is {\sl independent} of the height of the turbulence.

Absolute radiometric phase correction entails measuring variations in
brightness temperature with a radiometer, inverting equation 10 to
derive the variation in PWV, and then using equation 1 to derive the
variation in electronic phase along a given line of sight. 

As benchmark numbers for the MMA we set the requirement that we
need to measure changes in tropospheric induced phase above a given
antenna to an accuracy of  ${\lambda}\over{20}$ at 230 GHz at the
zenith, or $\phi_{\rm rms}$ = 18$^o$. This
requirement inserted into  equation 10 then yields a required
accuracy of:~ w$_{\rm rms}$ = 0.01 mm.
This value of w$_{\rm rms}$ then sets the required sensitivity,
T$_{\rm B}^{\rm rms}$, of the
radiometers as a function of frequency through equation 10. For the VLA
we  set the ${\lambda}\over{20}$
requirement  at 43 GHz, leading to:~ w$_{\rm rms}$ = 0.05 mm.
In its purest form, the  inversion of equation 10 requires:~ (i) a
sensitive, absolutely calibrated  radiometer, (ii) accurate
models for the run of temperature and pressure as a function of
height in the atmosphere, and  (iii) an accurate value for the height
of the PWV fluctuations.

Figure 10 shows the required sensitivity of the radiometer,
T$_{\rm B}^{\rm rms}$, given the benchmark numbers for w$_{\rm rms}$ for the VLA and
the MMA and using equation 10. It is important to keep in mind that
lower numbers on this plot imply that more sensitive radiometry is required
in order to measure the benchmark value of w$_{\rm rms}$.
The required T$_{\rm B}^{\rm rms}$ values generally
increase with increasing frequency due to the increase in A$_{\nu}$,
with a local maximum at the 22 GHz water line,
and minima at the strong O$_2$ lines (59.2 GHz and 118.8 GHz).
The strong water line at 183.3 GHz shows
a `double peak' profile, with a local minimum in T$_{\rm B}^{\rm rms}$ at the
frequency corresponding to the peak  of the line. This behavior
is due to the product:~ A$_{\nu}$ $\times$ e$^{-\tau_{\circ}}$ in equation 10.
The value of  A$_{\nu}$ peaks at the line frequency, but this is
off-set by the high total optical depth at the line peak.
This effect is dramatic for the VLA case,
where the required T$_{\rm B}^{\rm rms}$ at the 183 GHz line peak is very low.

\subsection{Absolute Radiometric Phase Corrections}

In this section we consider making an absolute correction to the
electronic phase at a given antenna using an accurate, absolutely
calibrated measurement of T$_{\rm B}$,
and accurate measurements of tropospheric parameters (temperature and
pressure as a function of height, and the scale height of the PWV
fluctuations). We consider requirements on the gain
stability, sensitivity, and on atmospheric data, given the benchmark
values of w$_{\rm rms}$ and
using equation 10 to relate w$_{\rm rms}$ and T$_{\rm B}^{\rm rms}$.
We consider the requirements at a number of frequencies for the MMA
site, including: (i)  the water lines at 22.2 GHz and 183.3 GHz, (ii)
the half power of the water line at 185.5 GHz, and (iii) two continuum
bands at 90 GHz and 230 GHz. For the VLA we only consider the 22.2
GHz line.

The results are summarized in Table 1. Row 1 shows the optical 
depth per mm PWV, A$_\nu$, at the different frequencies for the model
atmospheres discussed in section 4, while row 2 shows the total
optical depth, $\tau_{\rm tot}$, for the models. Row 3 shows the required
T$_{\rm B}^{\rm rms}$ values as derived from equation 10. It is important to
keep in mind that these values are simply the expected change in 
T$_{\rm B}$ given a change in w of 0.01 mm for the MMA and 0.05 mm for the
VLA, for a single radiometer looking at the zenith. All subsequent
calculations depend on these basic T$_{\rm B}^{\rm rms}$ 
values. The values range from 19 mK at 90 GHz,
to 920 mK at  185.5 GHz,  at the MMA site, and 120 mK for the VLA site
at 22 GHz. 

We first consider sensitivity and gain stability. Row 4 lists 
approximate numbers for expected receiver temperatures,
T$_{\rm rec+spill}$,  in the case of 
cooled systems (eg. using the astronomical receivers for radiometry).  
Row 5 lists the contribution to the system temperature from the
atmosphere, T$_{\rm rec,atm}$, and row 6 lists the 
expected total system temperature, T$_{\rm tot}$ (sum of row 4 and 5).
Row 7  lists the rms sensitivity of the radiometers, T$_{\rm rms}$, assuming
1000 MHz bandwidth, one polarization, and a 1 sec
integration time. In all cases the expected sensitivities of the
radiometers are well
below the required T$_{\rm B}^{\rm rms}$ values, indicating that sensitivity
should not be a limiting factor for these systems. Row 8 lists the
required gain stability of the system, defined as the ratio of total system
temperature to T$_{\rm B}^{\rm rms}$:~ $\delta$Gain $\equiv$
$\rm {T_{\rm tot}}\over{T_B^{\rm rms}}$. Values range from 210 for the
185.5 GHz 
measurement to 5800 for the 90 GHz measurement at the MMA, and 450 for
the VLA site at 22 GHz.

Rows 9 and 10 list total system temperatures and expected rms
sensitivities in the case of uncooled  radiometers.
We adopt a constant total system temperature of
T$_{\rm tot}$ = 2000 K, but the other parameters remain the same
(bandwidth, etc...). The radiometer sensitivity is then 63 mK in 1 second.
This sensitivity is adequate
to reach the benchmark T$_{\rm B}^{\rm rms}$ values in row 3, although at 230
GHz the sensitivity value is within a factor two of the required
T$_{\rm B}^{\rm rms}$.  The required gain stabilities in this case are
listed in 
row 11.  The requirement becomes severe at 230 GHz
($\delta$Gain = 15000). 

We next consider the requirements on atmospheric data, beginning with
T$_{\rm atm}$. The dependence of T$_{\rm B}^{\rm rms}$ on T$_{\rm atm}$ comes in
explicitly in equation 10 through the first multiplier, and implicitly
through the effect of T$_{\rm atm}$ on $\tau_{\circ}$. For simplicity, we
consider only the explicit dependence, 
which will lead to an under-estimate of the 
expected errors by at most a factor $\approx$ 2 -- adequate for the 
purposes of this document (Sutton and 
Hueckstaedt 1997).  Under this simplifying assumption the required accuracy,
$\delta$T$_{\rm atm}$, becomes: 
$$\rm \delta T_{\rm atm} ~ \approx ~ {{T_B^{\rm
rms}}\over{[1-e^{-\tau_{\circ}}]}} ~ K$$ 
The values $\delta$T$_{\rm atm}$ are listed in row 12. Values in
parentheses are the percentage accuracy in terms of the ground
atmospheric temperature. Values are typically of order 1 K, or a few
tenths of a percent 
of the mean. A related requirement is the accuracy of the gradient in
temperature:~ $\delta$$\rm {dT}\over{dh}$ $\approx$ $\rm {\delta
T_{\rm atm}}\over{h_{\rm turb}}$, ~ in 
the case of a turbulent layer at h$_{\rm turb}$ = 2 km, and
assuming a very accurate measurement of T$_{\rm atm}$ on the ground
and a very accurate measurement of h$_{\rm turb}$.
These values are listed in row 13. 
The accuracy requirements range from 0.25 K km$^{-1}$ to
1.5 K km$^{-1}$, or roughly 10$\%$ of the mean gradient. 
Similarly, we can consider the required accuracy of the measurement of
the height of the troposphere, $\delta$h$_{\rm turb}$ 
$\approx$ $\rm {\delta T_{\rm atm}}\over{{{dT}\over{dh}}}$,
assuming a perfect measurement of the ground
temperature and temperature gradient.
These values are listed in row 14. Values are typically a few tenths
of a km, or roughly 10$\%$ of h$_{\rm turb}$. 

We consider the requirements on atmospheric pressure given
the T$_{\rm B}^{\rm rms}$ requirements. The relationship between T$_{\rm B}$ and w
is affected by atmospheric pressure through the change in the pressure
broadened line shapes. An increase in pressure will transfer power
from the line peak into the line wings, thereby flattening the overall
profile. The expected changes in optical depth (or brightness
temperature) as a function of frequency have been quantified by 
Sutton and Hueckstaedt (1997), and their coefficients relating changes
in pressure with changes in optical depth are listed in row 17. Note
the change in sign of the coefficient on the line peaks versus
off-line frequencies. Sutton and Hueckstaedt point out that, since
the integrated power in the line is conserved, there
are `hinge points' in the line profiles where pressure changes have
very little effect on T$_{\rm B}$, ie. for an increase in pressure  at fixed
total PWV the wings of the line get broader while peak gets lower.
These hinge points are close
to the half power points in T$_{\rm B}$ of the lines. 
Rows 15 and 16 list the requirements on the accuracy of P$_{\rm atm}$, 
and on the value of h$_{\rm turb}$.
The values of $\delta$P$_{\rm atm}$ are derived from the equation:~ 
$\delta$P$_{\rm atm}$ = $\rm {A_{\nu}w_{\rm rms}}\over{\tau_{\rm tot}X}$,
where $\rm X$ is the coefficient listed in  row 17.
We find that the value of P$_{\rm atm}$ needs to be known to about 1$\%$,
and the height of the turbulent
layer needs to be known to a few percent. The exception is
at the hinge point of the line ($\approx$ 185.5 GHz), where the
optical depth is nearly independent of P$_{\rm atm}$.  

There are a few potential difficulties with absolute radiometric phase
corrections which we have not considered. First
is the question of how to
make a proper measurement of the `ground temperature'?
It is possible, and perhaps
likely, that the expected linear temperature gradient 
of the troposphere displays a
significant perturbation close to the ground. The method for making
the `correct' ground temperature measurement remains an important
issue to address in the context of absolute radiometric phase
correction. Second, we have only considered a simple model in
which the PWV 
fluctuations occur in a narrow layer at some height h$_{\rm turb}$, which 
presumably remains constant over time.
If the fluctuations are distributed over a large range of altitude
then one needs to know the height of the dominant fluctuation at {\sl
each time} to convert T$_{\rm B}$ into electrical pathlength. And when
fluctuations at 
different altitudes contribute at the same time, this conversion
becomes problematical. Again, the required accuracies for the height of
the fluctuations are given in rows 14 and 16 in Table 1. 
A possible solution to this problem is to 
find a linear combination of channels for which
the effective conversion factor is insensitive to altitude under a
range of conditions, ie. hinge points generalized to a multi-channel
approach (Lay 1998, Staguhn et al., 1998). 
And third, the shape of the pass band of the radiometer
needs to be known very accurately in order to obtain absolute T$_{\rm B}$
measurements.

A final uncertainty involved in making absolute radiometric phase
corrections are errors in the theoretical atmospheric 
models relating w and T$_{\rm B}$ (Elgered 1993). Sutton and
Hueckstaedt (1997)  point out that  model errors are by far the
dominant uncertainties  
when considering absolute radiometric phase correction, and they
have calculated a number of models with different line shapes and 
different empirically determined water vapor continuum
`fudge-factors'. Row 18 in Table 1 lists the approximate differences
between the various  models at various frequencies. Models can differ
by up to 3 mm in PWV, corresponding to 19.5 mm in electrical
pathlength, or 30$\pi$ rad in electronic phase at 230 GHz.
The differences are most pronounced in the continuum bands, but
are only negligible close to the peak of the strong 183 GHz
line. This is an area of very active research, and it 
may be that this uncertainty is greatly reduced in the near future
(Rosenkranz 1998). 

Given the status of current atmospheric models,  radiometric 
phase correction then requires some form of empirical
calibration of the water vapor continuum contribution in order to 
relate T$_{\rm B}$ to w. The exception may be a measurement close
to the peak of the 183 GHz line, but in this case saturation 
becomes a problem.  Perhaps most importantly, 
if the calibrated continuum term is due to incorrect line shapes, it
will depend on both T$_{\rm atm}$ and P$_{\rm atm}$, in which case 
the continuum term may require frequent calibration.

Overall, absolute radiometric phase correction requires: (i)  systems that
are sensitive (19 $\le$ T$_{\rm rms}$ $\le$ 920 mK), and 
stable over long timescales (200 $\le$ $\delta$Gain $\le$ 15000), and (ii) 
knowledge of the tropospheric parameters, such as
T$_{\rm atm}$, P$_{\rm atm}$, and h$_{\rm turb}$, to a few percent or less. 
And even if such accurate measurements are available, fundamental
uncertainties in the atmospheric models relating T$_{\rm B}$ and PWV may 
require empirical calibration of the T$_{\rm B}^{\rm rms}$ - w$_{\rm rms}$
relationship at regular intervals.

Lay (1998) has recently presented an interesting radiometric phase
correction method using
multifrequency measurements of the 183 GHz water line profile.
His method is insensitive to  atmospheric
parameters, since it relies on using the line profile, and in
particular, the hinge points of the lines. This method may
allow for an absolute radiometric phase correction to be made
without great uncertainties due to the atmospheric models.

\subsection{Empirically Calibrated Radiometric Phase Corrections}

Many of the uncertainties in Table 1 arise from the fact that we are
demanding  an absolute phase correction at each antenna
based on the measured T$_{\rm B}$
plus ancillary data (T$_{\rm atm}$, P$_{\rm atm}$, h$_{\rm turb}$,...), using a
theoretical model of the atmosphere to relate  T$_{\rm B}$ to w. This sets
very  stringent demands on  the absolute calibration, on the
accuracy of the ancillary data, and on the accuracy of the 
theoretical model atmosphere. The current atmospheric 
models under-predict w by large factors in the continuum bands,
thereby requiring calibration of (possibly time dependent) 
water vapor continuum `fudge-factors'. 

One way to  avoid some of these problems is to calibrate
the relationship between fluctuations in 
T$_{\rm B}^{\rm rms}$ with fluctuations in antenna-based phase,
$\phi_{\rm rms}$, by observing a strong celestial calibrator at regular
intervals. This empirically  calibrated phase correction method 
would circumvent dependence on ancillary data
and model errors (Woody and Marvel 1998), and mitigate long term gain 
stability problems in the electronics. This technique can be thought
of as calibrating the `gain' of both the atmosphere and the
electronics, in terms of  relating T$_{\rm B}^{\rm rms}$ to
$\phi_{\rm rms}$. 

In its simplest form, empirically calibrated radiometric phase
correction would be used only to increase the coherence time on
source. No attempt would be made to connect the phase of a celestial
calibrator with that of the target source using radiometry, 
and hence the absolute phase on the target source would still be
obtained from the calibration source.  
Such a process is being implemented at the
Owen Valley Radio Observatory (Woody and Marvel  1998). 
In this case the absolute phase is obtained from the first accurate
phase measurement on
the celestial calibrator, while the subsequent time series of phase 
measurements on the calibrator are then used to derive the 
T$_{\rm B}^{\rm rms}$ to $\phi_{\rm rms}$ relationship. 
This process results in additional 
phase uncertainty in a manner analogous to 
Fast Switching phase calibration (Holdaway and Owen 1995).
The residual error is set by the distance between the calibrator and
source, and the time required to obtain the first accurate record:
t$_{cal} \equiv$ (the slew time + the integration time
required for the first accurate phase measurement). In this case:
$$\rm b_{eff}~ \approx~ v_{a} \times t_{cal}~ + ~d,$$
where d is the physical distance in the troposphere set by
the angular separation of the calibrator and the source, and
b$_{eff}$ is the `effective baseline' to be inserted into equation 3
in order to estimate the residual uncertainty in the absolute phase.
For example, assuming t$_{cal}$ = 10 sec, and
the calibrator-source separation = 2$^o$,
leads to b$_{eff}$ = 170 m, or
$\phi_{\rm rms}$ = 20$^o$ at 230 GHz at the MMA site. Note that the
temporal character of this  `phase noise' is unusual in that the short
timescale (t $<<$ t$_{cyc}$) variations are removed by radiometry,
while the long timescale variations (t $>>$ t$_{cyc}$) are removed by
celestial source calibration (Lay 1997).  

An example of such a relatively calibrated radiometric phase
correction is shown in Figure 11, using data from the VLA at 22 GHz. 
Observations were made 
of the celestial calibrator 0319+415 (3C 84).
In the upper frame, 
the dash line shows the interferometric phase time series measured
between antennas 5 and 9, corresponding  to a baseline length of about
3 km. The solid line shows the predicted phase time series derived by 
differencing measurements of the 22 GHz system temperature  at each
antenna. A single scale factor relating phase
fluctuations  and fluctuations in system temperature differences was
derived  from all the data, by requiring a minimum residual rms
scatter in the phase fluctuations after applying radiometric phase
correction. A constant off-set was also applied to each data set. 
Note the clear correlation between measured interferometric phase
variations, and the phase variations predicted by radiometry. 
The middle frame shows the residual phase variations after radiometric
correction. The rms variations in the raw phase time series before
correction are  32$^o$. After applying the radiometric correction, the
rms phase variations are reduced to 17$^o$. 
The lower frame shows the residual phase variations after radiometric
correction, but now making a correction for the first and second
half of the data separately. The residual rms phase variations are now
13$^o$. The scale factor changes by about 10$\%$ over the 36
minutes.  

This `empirically calibrated' radiometric phase correction  technique
has been implemented successfully at the Owens Valley Radio 
Observatory and at the IRAM interferometer (Woody and Marvel 1998, 
Bremer et al., 1997).  
A number of questions remain to be answered concerning this
technique, including:
(i) over what time scale and distance will this technique  allow
for radiometric phase corrections when switching between
the source and the calibrator? And
(ii) how often will calibration of the T$_{\rm B}^{\rm rms}$ -
$\phi_{rms}$ relationship be required, ie. how stable are the
radiometers and the mean parameters of the atmosphere?

\subsection{Clouds and Other Issues}

Water droplets present the problem that the drops
contribute significantly to the measured T$_{\rm B}$ but not to w, thereby
invalidating the model relating T$_{\rm B}$ and w. This problem can be
avoided by using multichannel measurements around the water lines (183
GHz or 22 GHz), since T$_{\rm B}$ for the lines is not affected by water
drops. Alternatively, a dual-band system could be used to separate
the effect of water drops from water vapor (eg. 90 GHz and 230 GHz),
since the frequency dependence of T$_{\rm B}$ is different for the two
water phases. This later method requires a multi-band
radiometer, which may be difficult within the context of the MMA
antenna design. The question of whether clouds will be a significant 
problem  on high quality sites such as the MMA site in
Chile remains to be answered.

A final error introduced when using radiometric phase calibration is
due to the different path through the troposphere
seen by the radiometer with that seen by the
astronomical receiver. If the radiometer is not the astronomical
receiver itself, then the angular separation of the radiometer beam
and the telescope beam, $\theta_{diff}$, 
can be of order a few degrees, depending on the lay-out of the
receivers and the telescope optics.
This angle corresponds to 100 m or so at a height of 2 km. The
magnitude of the error introduced by such an observing path difference
can be derived from the root phase structure function assuming
b$_{eff}$ = h$_{\rm turb}$ $\times$ $\theta_{diff}$. 

\section{Fast Switching vs. Paired Array Calibration vs. Radiometric
Phase Correction}

Fast Switching phase calibration has been used extensively, and
successfully,  at the VLA for diffraction limited imaging  of faint
astronomical sources on 30 km baselines at 43 GHz
(Lim et al., 1998, Wilner et al., 1996).
However, the required switching times at higher frequencies (10's of
seconds or less) has thus
far precluded the use of FS for existing mm observatories. 
Paired Array calibration has been 
demonstrated effective in reducing tropospheric phase noise for 
interferometers (Asaki et al., 1996, 1998),
but has not yet been implemented as a standard technique for 
astronomical observing.  Radiometric phase correction has been used
to varying degrees of success at different observatories, but mostly
on an experimental basis (Welch 1994, Bremer 1997, Woody and Marvel
1998, Elgered et al., 1991).  We summarize the relative advantages and
disadvantages of the three techniques. 

The advantages of FS are that: (i) FS uses the full array to observe
the target source, and (ii) FS removes the long and short term
electronic phase noise along with the tropospheric phase noise. The
disadvantages are that: (i) FS places stringent constraints on
telescope design in terms of slew rate, mechanical settling time, and
electronic set-up time, and (ii) on-source observing time is lost due
to frequent moves and calibration.
Simulations for the MMA (Holdaway 1998)
indicate that overall fast switching efficiencies (including both
sensitivity losses due to decreased observing time and increased
decorrelation) of about 0.75 should be possible,
assuming a distribution of observing frequencies from 30~GHz to 650~GHz
and using the measured distribution of atmospheric phase conditions
at the Chajnantor site.  Typical switching times range from 10 to
20 seconds. 

The advantages of PA calibration are that: (i) the `target
array' observes the source continuously, and (ii) the demands on the
antenna mechanics and electronics are less stringent than for FS. The
disadvantages are that: (i) the electronic phase noise is not
removed, (ii) the geometry of the array must allow for neighboring
antennas, even in large (sparsely populated) arrays, and (iii) the
number of visibilities 
from the target source array decreases quadratically with the
decreasing number of antennas. These latter two effects can reduce
significantly the Fourier spacing coverage and sensitivity of the array. 
One possible solution to these problems 
is to have a `calibration array' of smaller,
cheaper antennas strategically placed with respect to the antennas of
the main array which 
is  dedicated to tropospheric phase calibration by observing celestial
calibrators  at a fixed frequency (eg. 90 GHz). 
The solutions would then be extrapolated to the
observing frequency of the main array using the linear relationship
between tropospheric phase and frequency (equation 1).
This would require a separate correlator and IF system, and
complications may arise due to the fact that the wet troposphere
becomes dispersive at  frequencies higher than about 
400 GHz (Sutton and Hueckstaedt 1997).

For PA and FS calibration, 
if the bulk of the phase fluctuations occur in a
thin turbulent layer, it may be possible to perform an 
intelligent method of data interpolation, such as forward 
projection using a physical model for the troposphere and
Kalman filtering of the spatial time series,
to account for the motion of the atmosphere across the array.  

The advantages of radiometric phase correction are that it: (i) 
alleviates constraints on telescope  
mechanics and array design, and (ii) the full array can observe
the target source continuously.  The disadvantages are that: (i) it
places constraints on receiver lay-out such that the radiometer is always
looking at the sky, (ii) it does not remove the electronic phase
noise, and (iii) there remains significant questions about the
viability of absolute radiometric phase correction, or the limitations 
imposed when using an  `empirically calibrated' radiometric phase
correction  technique (see sections 7.1 and  7.2).

All three calibration methods are being investigated for the MMA, and 
we envision that all the methods will be employed to some degree
at the MMA, depending on the configuration, the observing conditions,
and the scientific requirements of a given observation.

\footnotesize\acknowledgments

We thank  F. Owen,
O. Lay, E. Sutton, and J. Carlstrom for useful comments on sections 
of  this paper, and K. Desai 
for allowing us to use a figure from a previous publication.
This research made use of the NASA/IPAC Extragalactic Data Base (NED)
which is operated by the Jet propulsion Lab, Caltech, under contract
with NASA. The National Radio Astronomy Observatory is a facility of
the National Science Foundation operated under cooperative 
agreement by Associated Universities, Inc.


\clearpage
\newpage

\begin{planotable}{llllllll}
\tablewidth{44pc}
\footnotesize
\tablecaption{Limits to Absolute Radiometric Phase Calibration}
\tablenum{1}
\tablehead{
\colhead{~} &
\colhead{~} &
\colhead{22.2 GHz} &
\colhead{90} & 
\colhead{183.3} &
\colhead{185.5} &
\colhead{230} &
\colhead{22.2 (VLA)}
}
\startdata
\label{tbl-1}
1 & A$_\nu$  ~ mm$^{-1}$ & 0.0115 & 0.0073 & 2.79 & 0.670 & 0.053 & 0.0085 \nl 
2 & $\tau_{\rm tot}$ & 0.0167 & 0.028 & 2.79 & 0.677 & 0.057 & 0.043 \nl
3 & T$_{\rm B}^{\rm rms}$ ~ mK & 30 & 19 & 460 & 920 & 135 & 120 \nl
4 & T$_{\rm rec+spill}$ ~ K & 40 & 100 & 100 & 100 & 100 & 40 \nl
5 & T$_{\rm rec,atm}$ ~ K & 7 & 10 & 246 & 94 & 20 & 14 \nl
6 & T$_{\rm tot}$ ~ K & 47 & 110 & 346 & 194 & 120 & 54 \nl
7 & T$_{\rm rms}^{\rm cool}$ ~ mK & 1.4 & 3.5 & 11 & 6 & 3.8 & 1.7 \nl
8 & $\delta$Gain & 1600 & 5800 & 750 & 210 & 890 & 450 \nl
9 & T$_{\rm tot}$ ~ K & -- & -- & 2000 & 2000 & 2000 & -- \nl
10 & T$_{\rm rms}^{uncool}$ ~ mK & -- & -- & 63 & 63 & 63 & -- \nl
11 & $\delta$Gain & -- & -- & 4400 & 2200 & 15000 & -- \nl
12 & $\delta$T$_{\rm atm}$ ~ K & 1.8 (0.7$\%$) & 0.7 (0.3$\%$) & 0.5 (0.2$\%$) &
1.9 (0.7$\%$) & 2.4 (0.9$\%$) & 2.9 (1.0$\%$) \nl
13 & $\delta$$\rm{d T_{\rm atm}}\over{d h}$ ~ K/km & 0.9 (14$\%$) & 0.35
(5$\%$) & 0.25 (4$\%$) & 1.0 (15$\%$) & 1.2 (18$\%$) & 1.5 (22$\%$)\nl 
14 & $\delta$h$_{\rm turb}$ ~ km & 0.28  (14$\%$) & 0.11  (5$\%$) & 0.08  (4$\%$)
& 0.29  (15$\%$) & 0.37  & 0.45 (22$\%$) \nl
15 & $\delta$P$_{\rm atm}$ ~ mb & -5 (0.9$\%$) & 2 (0.3$\%$) & -7.5 (1.3$\%$)
& $\infty$ & 7 (1.2$\%$) & -7.4 (1.0$\%$) \nl
16 & $\delta$h$_{\rm turb}$ ~ km &  0.07 (3$\%$) & 0.03 (1.5$\%$)
& 0.1 (5$\%$) & $\infty$ & 0.1 (5$\%$) & 0.1 (5$\%$) \nl
17 & ${1}\over{\tau_{\rm tot}}$ ${d \tau}\over{d P}$  ~ mb$^{-1}$ &
-0.00133 & 0.00133 & -0.00133 & 0 & 0.00133 & -0.00133 \nl
18 & $\delta$(Model w)  mm & 0.2 & 3 & 0.0025 & 0.01 & 3 & 4.8 \nl
\enddata
\end{planotable}

\clearpage
\newpage

\centerline{\bf Notes to Table 1}

{\bf Basic Assumptions:} 
$\phi_{\rm rms}$ = 18$^o$ (${\lambda}\over{20}$) at 230 GHz
(MMA) and 43 GHz (VLA), and w$_{o}$ = 1mm (MMA) and  4mm (VLA)

{\bf A$_\nu$:} Optical depth per mm of water.

{\bf $\tau_{\rm tot}$:} Total optical depth of the model (H$_2$O  plus
dry air).

{\bf T$_{\rm B}^{\rm rms}$:} Required rms of the measured 
brightness temperature to meet the
${\lambda}\over{20}$ standards given above.

{\bf T$_{\rm rec+spill}$:} Total system temperature excluding the atmospheric
contribution.

{\bf T$_{\rm rec,atm}$:} Atmospheric contribution to the system temperature.

{\bf T$_{\rm tot}$:} Total system temperature on sky. Two different
assumptions are made at a few frequencies, for cooled and uncooled
systems. Row 6 gives the case of a cooled receiver
(eg. using the 
astronomical receivers for radiometry). Row 9  gives the case of an
uncooled receiver.

{\bf T$_{\rm rms}$:} Expected rms noise in 1 sec with 1 GHz bandwidth and 1
polarization.

{\bf $\delta$Gain:} Required gain stability in order to obtain T$_{\rm B}^{\rm rms}$.

{\bf $\delta$T$_{\rm atm}$:} Required accuracy of the measurement of the
atmospheric temperature (in the turbulent  layer) in order to
obtain T$_{\rm B}^{\rm rms}$. Values in parentheses indicate
the percentage of the total.

{\bf $\delta$$\rm {d T_{\rm atm}}\over{d h}$:} Required accuracy of the
measurement of the gradient in atmospheric temperature, assuming 
T$_{\rm atm}$ is measured very accurately on the ground and 
extrapolated to h$_{\rm turb}$.

{\bf $\delta$h$_{\rm turb}$:} Required accuracy of the measurement of the
height of the turbulent  layer given  the $\delta$T$_{\rm atm}$
requirement.

{\bf $\delta$P$_{\rm atm}$:} Required accuracy of the
measurement of the atmospheric pressure (in the turbulent
layer) in order to obtain T$_{\rm B}^{\rm rms}$.


%

{\bf $\delta$h$_{\rm turb}$:} Required accuracy of the measurement of the
height of the turbulent layer given the $\delta$P$_{\rm atm}$
requirement.

{\bf ${1}\over{\tau}$ ${d \tau}\over{d P}$:} Constants used to relate changes
in pressure to changes in optical depth as a function of frequency 
(Sutton and Hueckstaedt 1997).

{\bf $\delta$(Model w):} Differences between various model
atmospheres involving different line shapes and different `water vapor
continuum fudge-factors' relating the measured T$_{\rm B}$ to w
(Sutton and Hueckstaedt 1997).

\vfill\eject

\begin{figure}
\psfig{figure=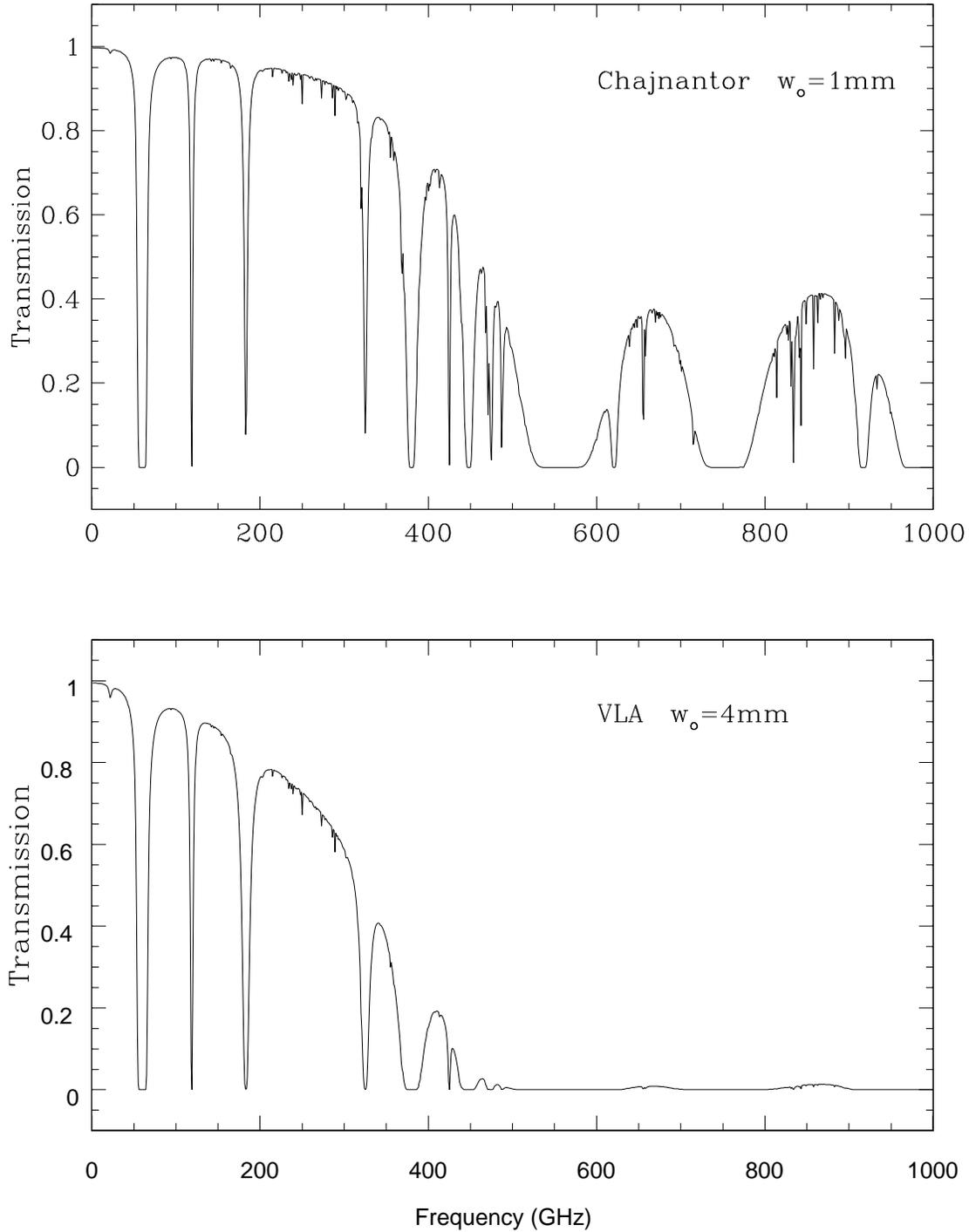,width=6in}
\caption{The upper frame shows the transmission of the atmosphere from
0 to 1000 GHz for the MMA site at Chajnantor in Chile assuming the
typical value of w$_{\circ}$ = 1 mm of precipitable water vapor, calculated
using the Liebe atmospheric model (Liebe 1989, Holdaway and Pardo
1997). The lower frame shows the transmission of the atmosphere for
the VLA site in Socorro, NM, assuming a value of w$_{\circ}$ = 4
mm.  
} 
\end{figure}
\vfill\eject

\begin{figure}
\psfig{figure=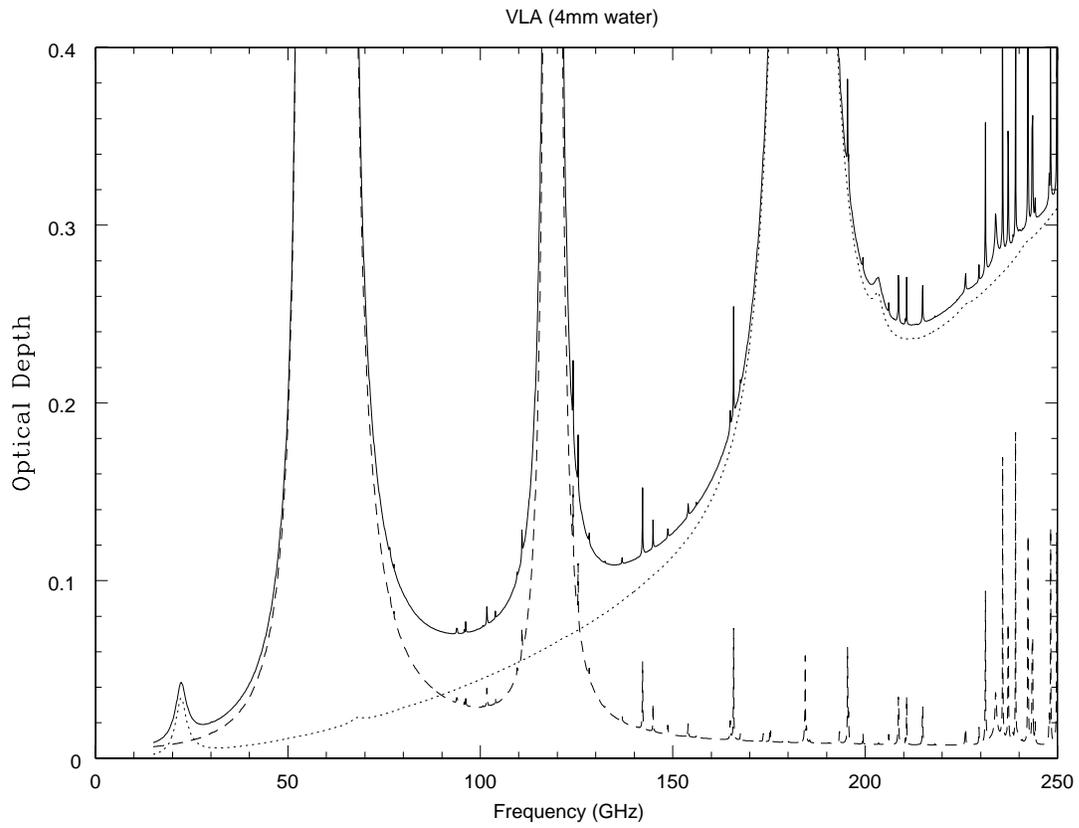,width=6in,angle=-90}
\caption{The optical depth for the VLA site assuming
w$_{\circ}$ = 4 mm. The 
solid line is the total optical depth. The dotted line is the optical
depth due to water vapor. The dash line is the optical depth due to
dry air (O$_2$  and other trace gases).}
\end{figure}
\vfill\eject

\begin{figure}
\psfig{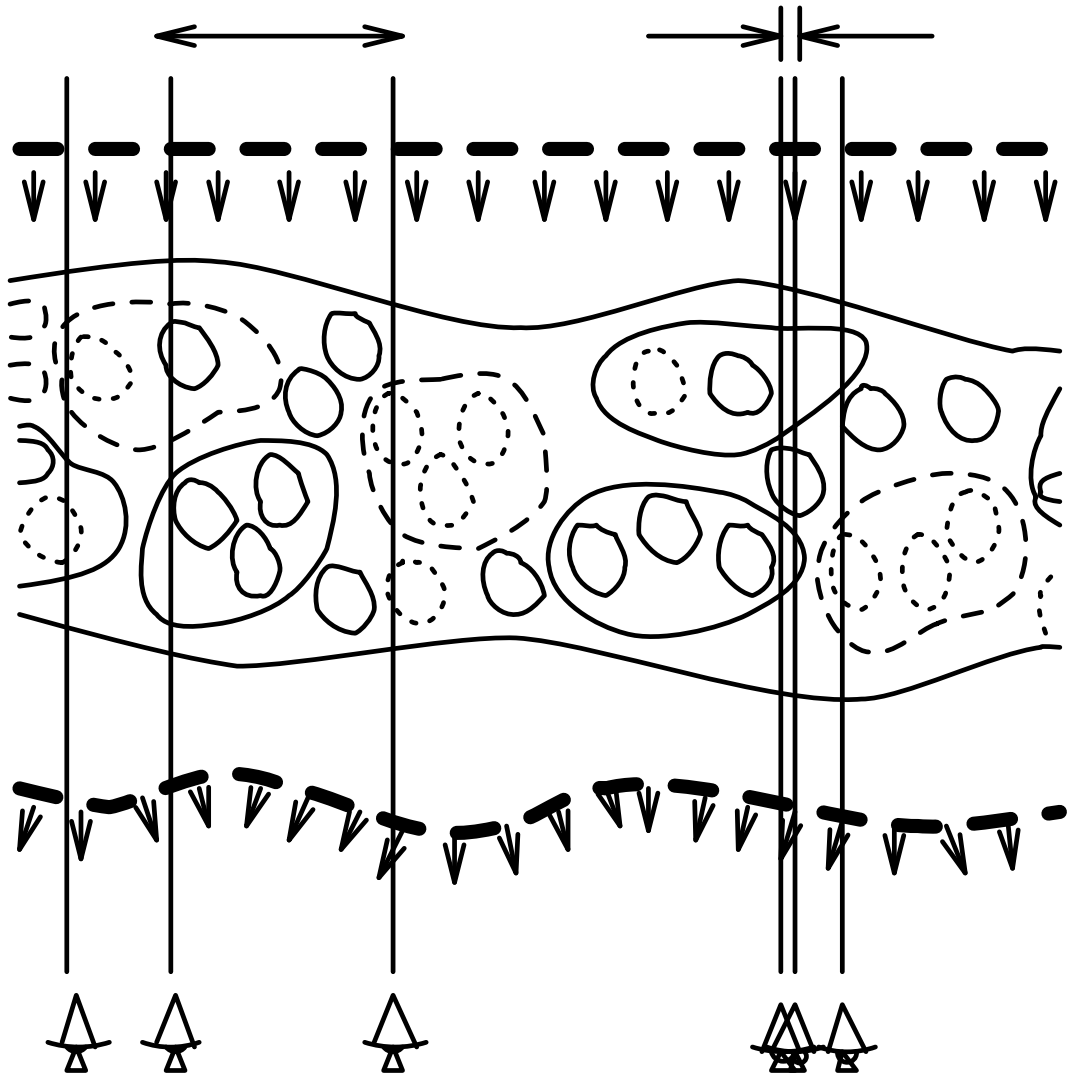}
\caption{A schematic diagram of the effect of structure in the
water vapor content of the atmosphere on different
scales on interferometric phase. Circles of different sizes represent
fluctuations in the water vapor content of the troposphere on various
scales, including excesses (solid contours) and deficits (dotted
contours). The phase of the incoming plane wave is distorted by the
variations in the index of refraction due to variations in the water
vapor content of the troposphere. The Taylor hypothesis implies that
this phase screen advects across the array with the mean 
velocity of the winds aloft.  Large scale fluctuations have the
largest amplitude, but the effect on interferometric phase 
for closely spaced antennas is
partially correlated. Smaller scale fluctuations have smaller
amplitude, but are not correlated between antennas (figure from 
K. Desai 1998). }
\end{figure}
\vfill\eject

\begin{figure}
\psfig{figure=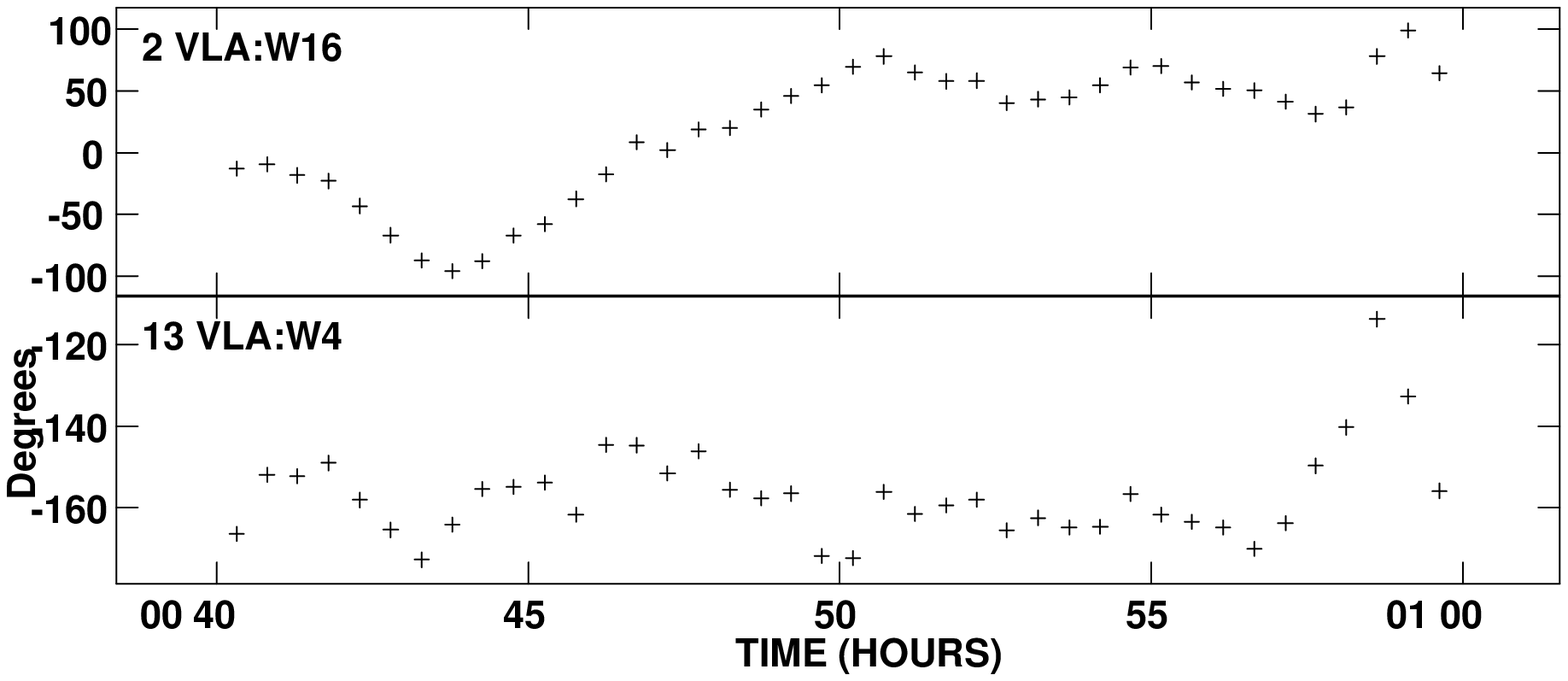,width=6in}
\vspace*{-3in}
\psfig{figure=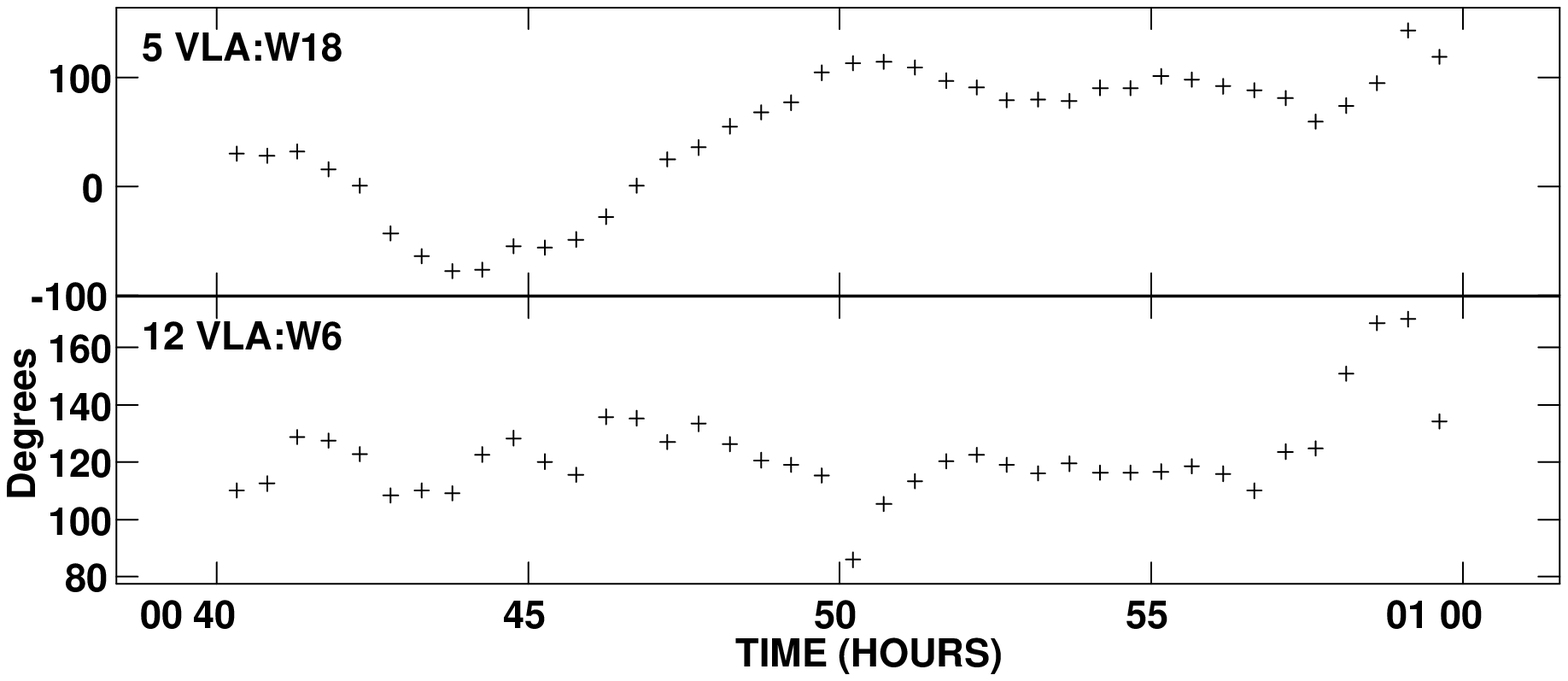,width=6in}
\vspace*{-3in}
\caption{The top figure shows the antenna-based phase solutions
vs. time for two antennas along the west arm of the VLA in a subarray
derived from observations of the 1 Jy 
celestial calibrator 0423+418 at 22 GHz on April 8, 1996.
The bottom figure shows 
the phase solutions over the same time for two antennas in a second
subarray along the west arm observing the 1 Jy celestial calibrator
0432+416. Antennas 5 (at station W18) and 2 (at W16) are at adjacent
positions, and antennas 12 (at W6) and 13 (at W4) are adjacent. 
The baseline length between antennas W18 and W16 is 360 m while
that between W6 and W4 is 148 m. The baseline length between W6 
and the reference antenna is 260 m, while that between W18 and the
reference antenna is 1910 m.} 
\end{figure}
\vfill\eject

\begin{figure}
\psfig{figure=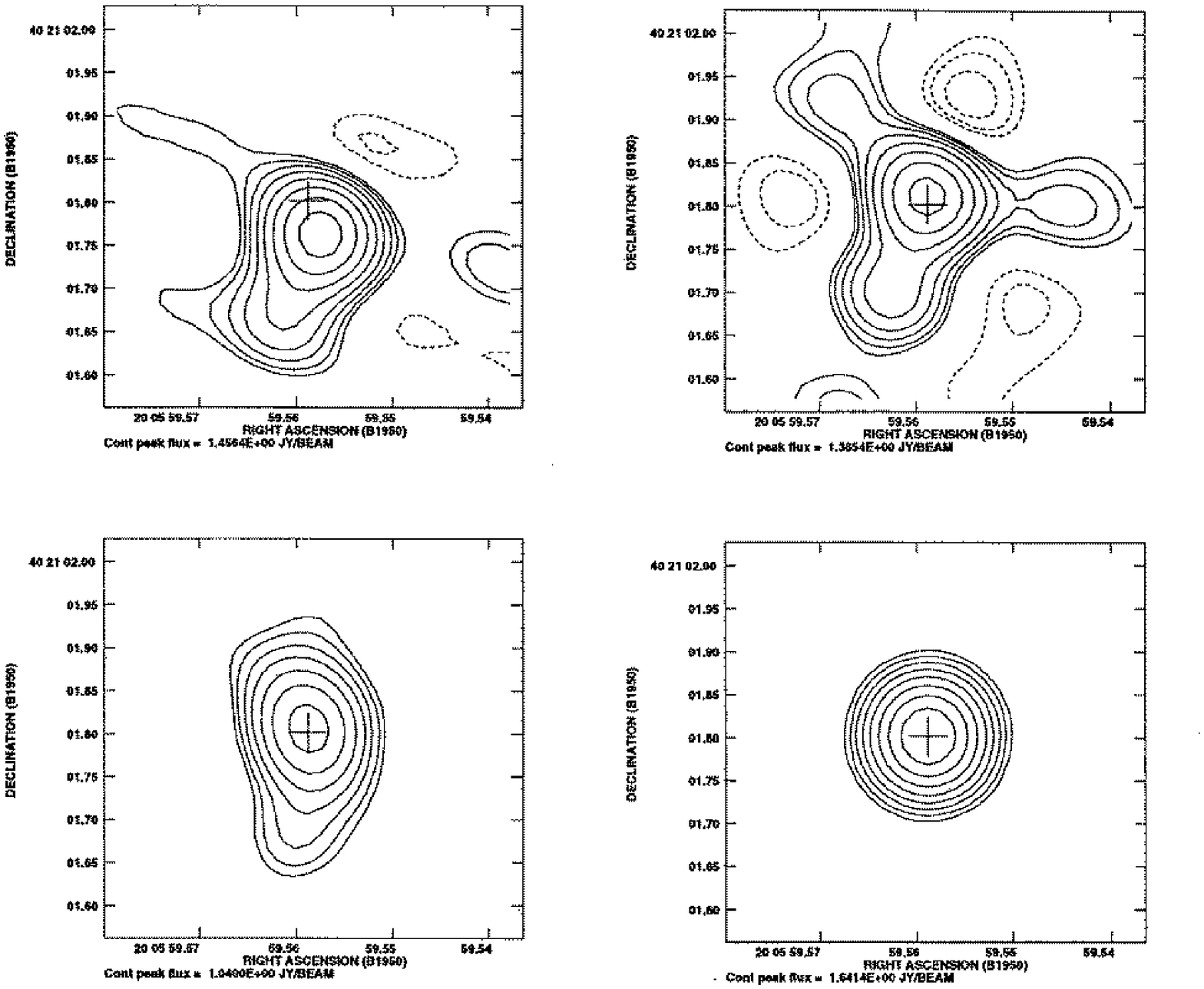,width=6in}
\caption{VLA images of the celestial calibrator 2007+404 at 22 GHz
with a resolution of 0.1$''$ made from observations on 
July 22, 1998. The tick-marks on the declination axis
are separated by 50 mas. The upper two frames show `snap-shot'
images made from one minute of data at the beginning (left) and end
(right) of the one hour observation. These data were self-calibrated
with an averaging time of 30 minutes, ie. just a mean phase was
removed for each half of the observation. The cross in each figure is
a fiducial mark indicating the true position of the source. The lower
left frame shows the image of 2007+404 made using the full hour of
data with a self-calibration averaging time of 30 minutes. The lower
right frame shows the same image after self-calibration with an
averaging time of 30 seconds. The lower left frame has a peak surface
brightness of 1.0 Jy beam$^{-1}$, and off-source rms noise of 47 mJy
beam$^{-1}$, and a total flux density of 1.5 Jy. The lower right frame
has a peak surface brightness of 1.6 Jy beam$^{-1}$, a noise of
5 mJy beam$^{-1}$, and a total flux density of 1.6 Jy. In all images
the contour levels are a geometric progression in the square root of
two, with the first level being 0.11 Jy beam$^{-1}$. Dotted contours
are negative.} 
\end{figure}
\vfill\eject

\begin{figure}
\psfig{figure=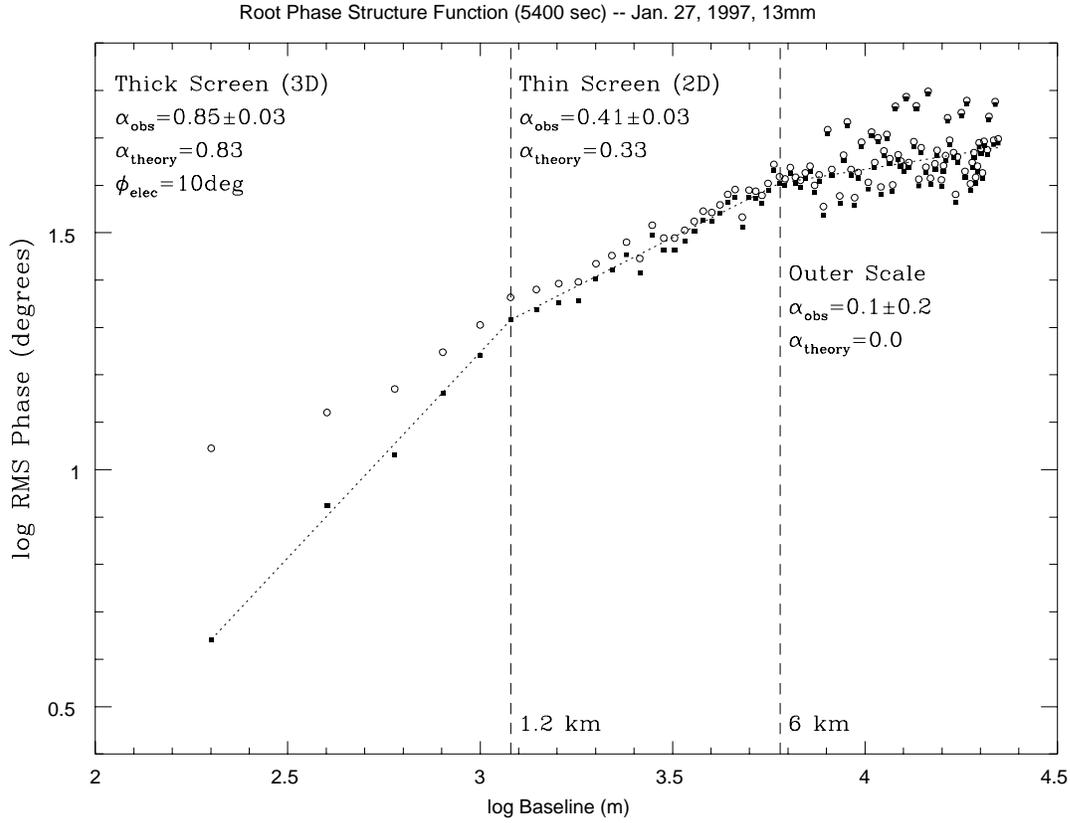,width=6in,angle=-90}
\caption{The root phase structure function from observations at 22 GHz
in the BnA configuration of the VLA on January 27, 1997 (Carilli and
Holdaway 1997). The open circles show the rms phase variations
vs. baseline length as measured on the 1 Jy celestial calibrator 0748+240
over a period of 90 minutes. The filled squares show these same values
with a constant noise term of 10$^o$ subtracted in quadrature. The
three regimes of the root phase structure function as predicted by
Kolomogorov turbulence theory are indicated. } 
\end{figure}
\vfill\eject

\begin{figure}
\psfig{figure=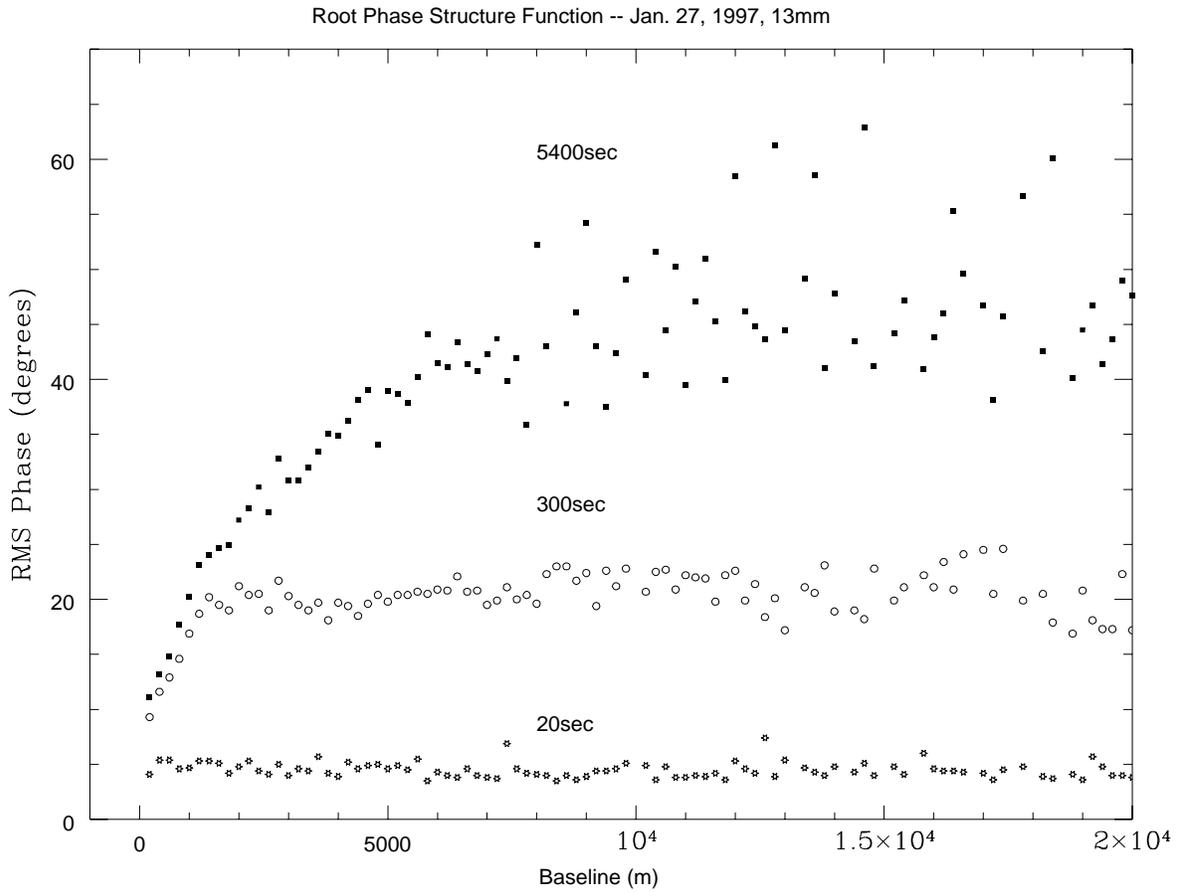,width=6.5in,angle=-90}
\caption{ The solid squares are the same as for Figure 6, but now on a
linear scale. These show the nominal 
root phase structure function from observations at 22 GHz
with the  VLA, as derived
from a 90 min phase time series on the celestial calibrator
0748+240. The open circles show the 
residual rms phase variations vs. baseline length after 
self-calibrating the data with an averaging time of 300 seconds. The
stars show the residual rms phase variations vs. baseline length after
calibrating with a cycle time of 20 seconds.} 
\end{figure}
\vfill\eject

\begin{figure}
\psfig{figure=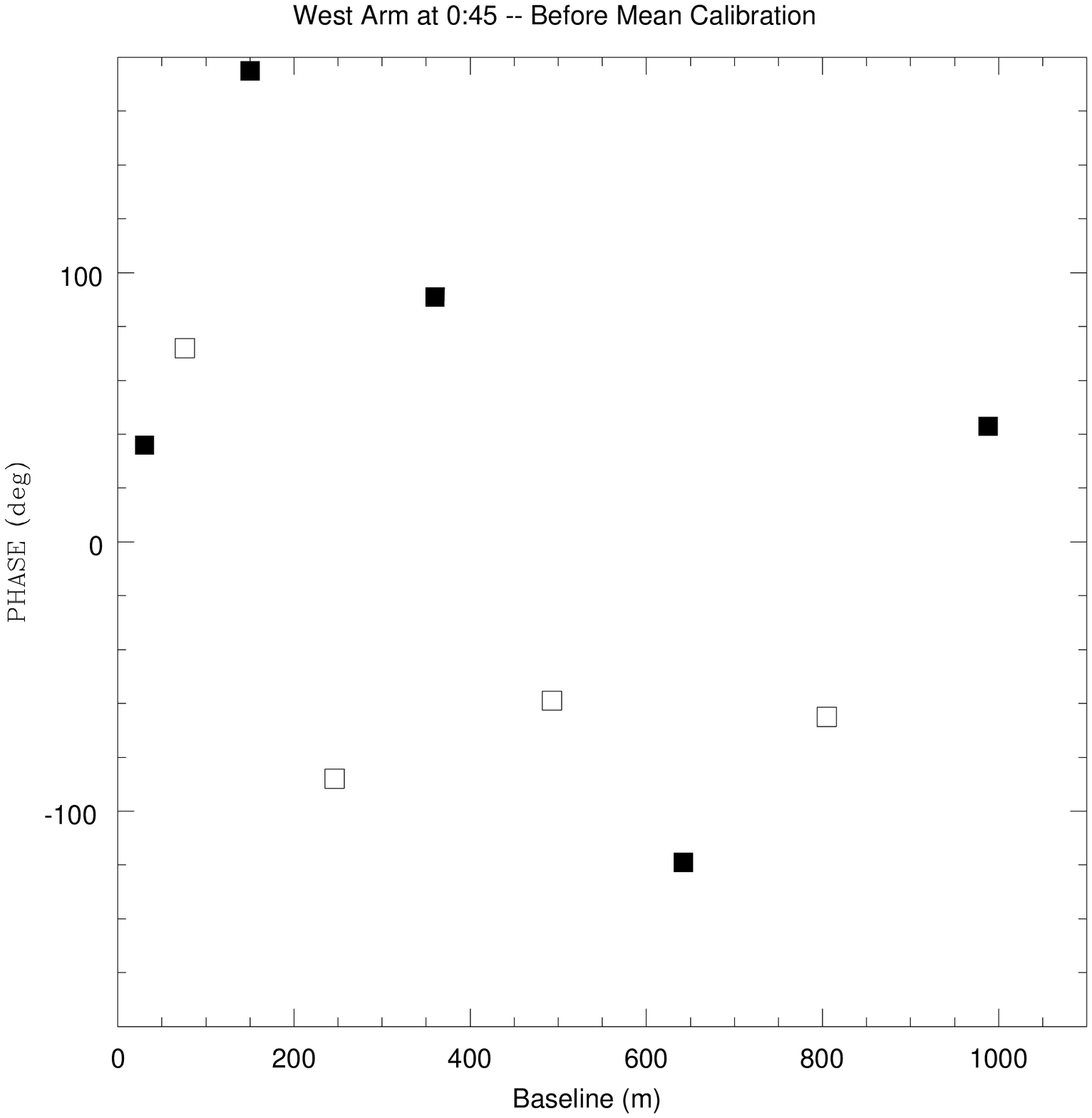,width=4in}
\psfig{figure=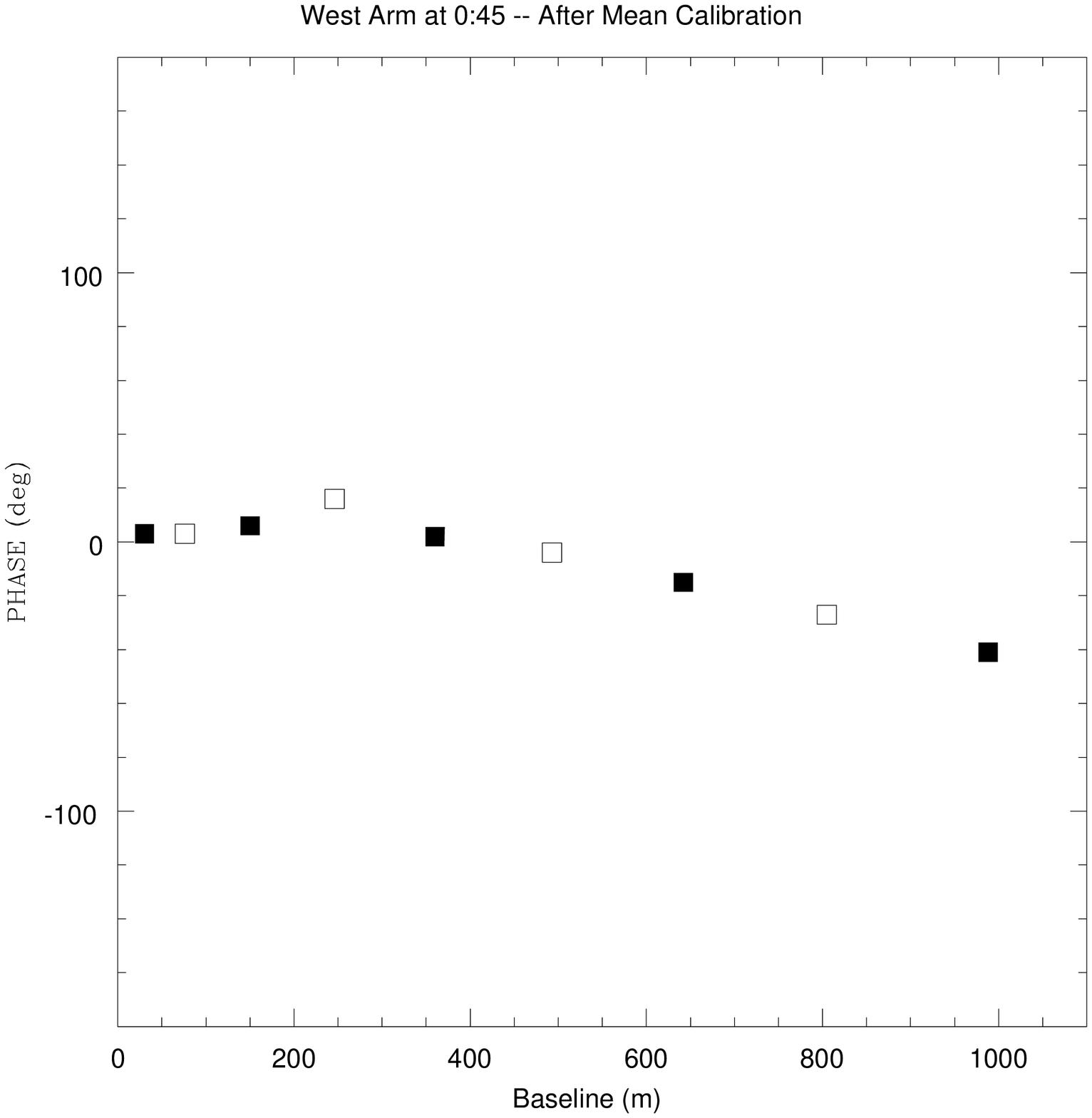,width=4in}
\caption{The top figure shows the antenna-based phase solutions 
at 22 GHz for antennas along the west arm of the VLA for a single 30
second observation (same data as in Figure 4). Two `inter-laced'
subarrays were employed. The 
solid squares are for antennas observing the celestial calibrator
0423+418 and the open squares are for antennas observing 
the celestial calibrator 0432+416. The bottom figure
shows the same  phase solutions after subtraction of the mean
electronic phase averaged over 20 minutes.
 } 
\end{figure}
\vfill\eject

\begin{figure}
\psfig{figure=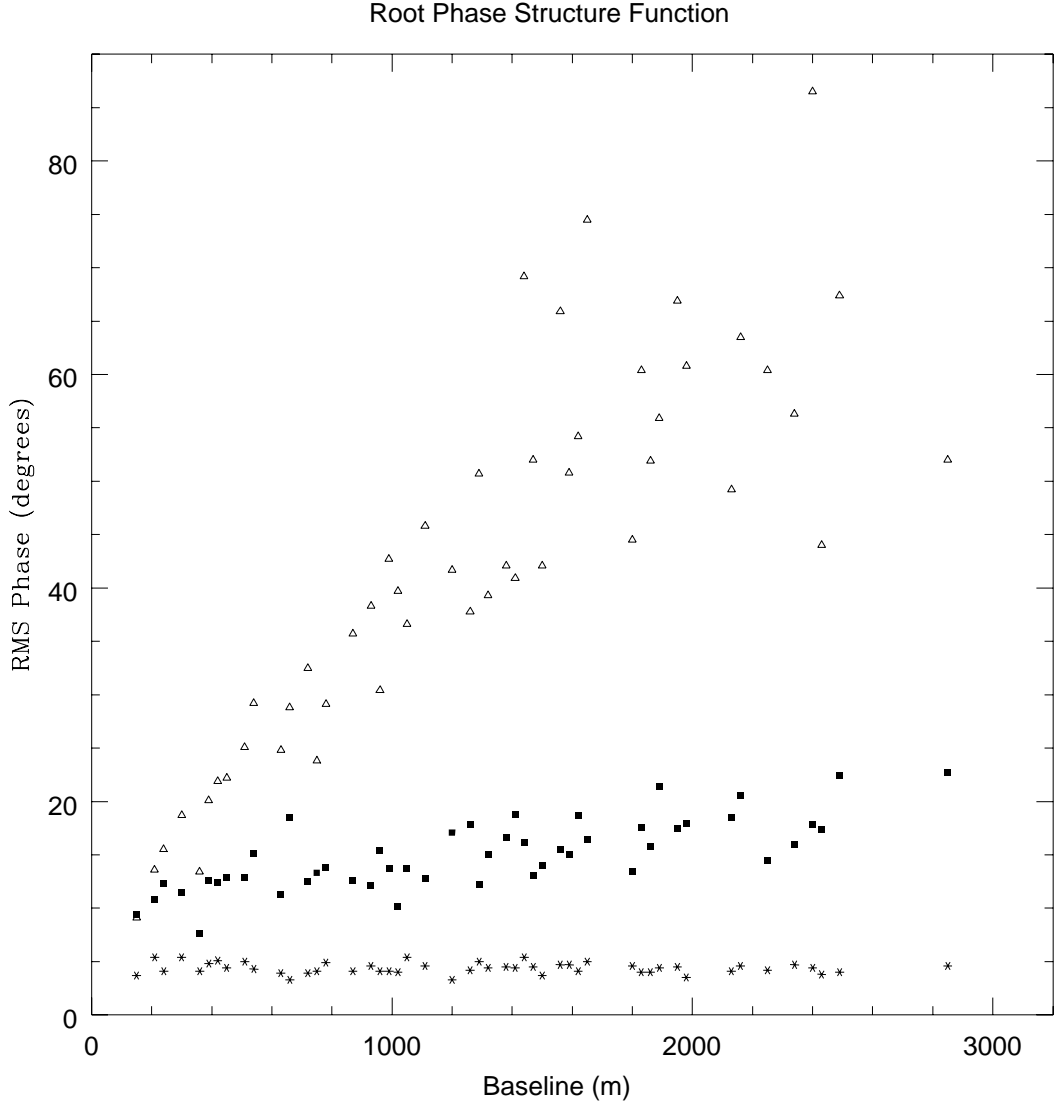,width=6in}
\caption{ The open triangles show the root phase structure function for
the data at 22 GHz on the `target source' 0432+416 (same data as
Figure 4) as determined from 
data with only a mean phase self-calibration applied (30 min
average). The stars show the structure function after application of
the self-calibration with 30 second averaging. The solid squares show
the structure function after application of Paired Array calibration
with a 30 second averaging time, which entails applying the phase
solutions from neighboring antennas observing the `calibration source'
0423+418. } 
\end{figure}
\vfill\eject

\begin{figure}
\psfig{figure=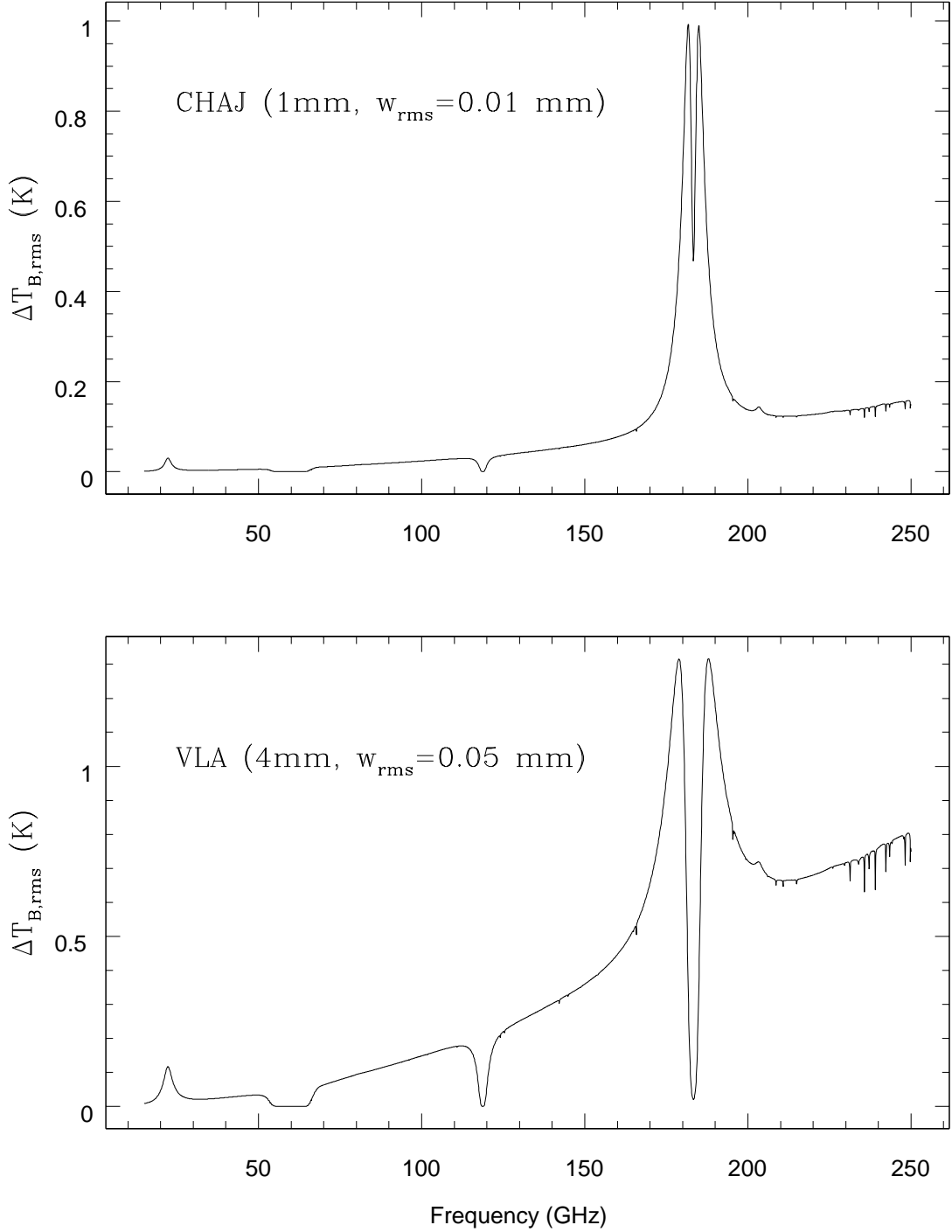,width=6in}
\caption{The upper frame shows the brightness temperature
sensitivities, T$_{\rm B}^{\rm rms}$, required to measure PWV
variations to an accuracy of w$_{\rm rms}$ = 0.01 mm, corresponding to
residual rms phase variations at 230 GHz of 
18$^o$, for the MMA site at Chajnantor assuming w$_{\circ}$ = 1 mm (equation
10). The bottom frame shows the  corresponding
T$_{\rm B}^{\rm rms}$ for the  VLA
site assuming w$_{\circ}$ = 4 mm and requiring w$_{\rm rms}$ =
0.05 mm, corresponding to residual rms phase variations at 43 GHz of
18$^o$. } 
\end{figure}
\vfill\eject

\begin{figure}
\vskip -0.6in
\psfig{figure=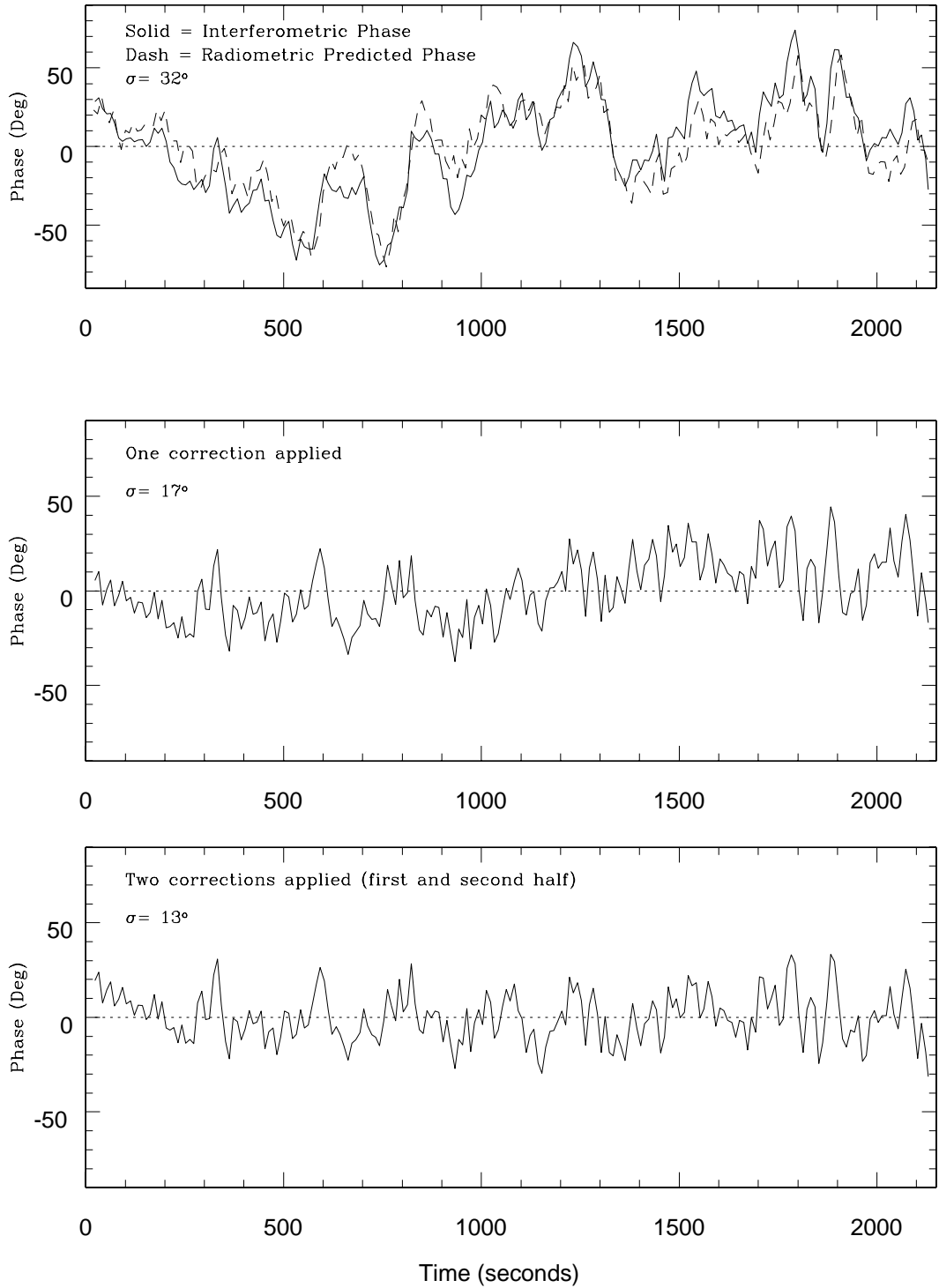,width=6in}
\vskip -0.1in
\caption{Upper frame: The dash line shows the interferometric phase
time series 
at 22 GHz  measured between VLA antennas 5 and 9 (baseline length = 3
km). The source observed was the 16 Jy  calibrator 0319+415 (3C 84)
on October 16, 1998.  
The solid line shows the predicted phase time series derived by 
differencing measurements of the 22 GHz system temperature 
at each antenna. The scale factor relating phase
fluctuations  and temperature fluctuations  was
derived  from all the data. The middle frame shows the residual phase
variations after radiometric phase correction using a single scale
factor derived from the entire time series. The lower frame
shows the same residuals, but now using corrections derived for the
first and second half of the data separately. 
The rms of the phase variations ($\sigma$) are given in each frame.} 
\end{figure}

\end{document}